\documentclass[journal]{IEEEtran}

\AtBeginDocument{%
  \providecommand\BibTeX{{%
    \normalfont B\kern-0.5em{\scshape i\kern-0.25em b}\kern-0.8em\TeX}}}

\usepackage[table,xcdraw]{xcolor}
\usepackage{array}
\usepackage{textcomp}
\usepackage{stfloats}
\usepackage{url}
\usepackage{verbatim}
\usepackage{graphicx}
\usepackage{pifont}
\usepackage{hhline}
\usepackage{adjustbox}
\usepackage{float}
\usepackage{booktabs}
\usepackage{multirow}
\usepackage{tabu}
\usepackage{lscape}
\usepackage[normalem]{ulem}
\usepackage{arydshln}
\usepackage{lettrine}
\usepackage{comment, soul}
\usepackage{bm}
\usepackage{subcaption}
\usepackage{diagbox}
\usepackage{slashbox}
\usepackage{makecell}
\usepackage{xspace}
\usepackage{acronym}
\usepackage{caption}
\usepackage[edges]{forest}
\usepackage{tikz}
\usetikzlibrary{shadows.blur}
\usepackage{paralist}
\usepackage{longtable}
\usepackage{threeparttable}

\usepackage{hyperref} 
\hypersetup{
    colorlinks=true,
    linkcolor=blue,
    filecolor=magenta,      
    urlcolor=cyan,
}

\usepackage{url}
\usepackage{tabularx,booktabs}
\usepackage{mwe,frame}
\usepackage{enumerate,multicol}
\usepackage{gensymb}
\usepackage{balance}
\usepackage{todonotes}
\usepackage{todonotes}
\usepackage{pgfplots}
\usetikzlibrary{spy,calc}

\usepackage{amsmath,amsfonts}
\usepackage{algpseudocode}
\usepackage{indentfirst}
\usepackage{ragged2e}

\usepackage[ruled,vlined]{algorithm2e}
\usepackage{wrapfig}
\usepackage{cite}

\usepackage{pgfplots}
\pgfplotsset{compat=1.18}
\newacro{lstm}[LSTM]{long short term memory}
\newacro{has}[HAS]{HTTP adaptive streaming}
\newacro{rcn}[RCN]{recurrent convolutional network}
\newacro{hevc}[HEVC]{high-efficiency video coding}
\newacro{vvc}[VVC]{versatile video coding}
\newacro{avc}[AVC]{advanced video coding}
\newacro{ott}[OTT]{over-the-top}
\newacro{ml}[ML]{machine learning}
\newacro{dl}[DL]{deep learning}
\newacro{rq}[RQ]{rate-quality}
\newacro{cu}[CU]{coding unit}
\newacro{pu}[PU]{prediction unit}
\newacro{mv}[MV]{motion vector}
\newacro{rd}[RD]{rate-distortion}
\newacro{pf}[PF]{Pareto front}
\newacro{fc}[FC]{fully connected}
\newacro{eel}[EEL]{exhaustive encoding ladder}
\newacro{pstr}[PSTR]{per-title encoding using spatio-temporal resolutions}
\newacro{lts}[LTS]{long term support}
\newacro{vod}[VOD]{video on demand}
\newacro{qp}[QP]{quantization parameter}
\newacro{qoe}[QoE]{quality of experience}
\newacro{bd-br}[BD-BR]{bjøntegaard delta bitrate}
\newacro{hls}[HLS]{HTTP live streaming}
\newacro{dash}[DASH]{dynamic adaptive streaming over HTTP}
\newacro{vmaf}[VMAF]{video multi-method assessment fusion}
\newacro{rnn}[RNN]{recurrent neural network}
\newacro{avsd}[AVSD]{adaptive video streaming dataset}
\newacro{ctu}[CTU]{coding tree unit}
\newacro{crf}[CRF]{constant rate factor}
\newacro{svm}[SVM]{support vector machine}
\newacro{jnd}[JND]{just noticeable difference}
\newacro{svr}[SVR]{support vector regression}
\newacro{gpr}[GPR]{Gaussian processes regression}
\newacro{rfr}[RFR]{random forest regression}
\newacro{rf}[RF]{random forest}
\newacro{gop}[GOP]{group of pictures}
\newacro{si}[SI]{spatial information}
\newacro{ti}[TI]{temporal information}
\newacro{cf}[CF]{colorfulness}
\newacro{br}[BR]{brightness}
\newacro{dnn}[DNN]{deep neural network}
\newacro{crgu}[Conv-GRUs]{convolutional gated recurrent units}
\newacro{map}[MAP]{maximum a posteriori}
\newacro{glcm}[GLCM]{gray level co-occurrence matrix }
\newacro{tc}[TC]{temporal coherence}
\newacro{ncc}[NCC]{normalized cross correlation}
\newacro{rfe}[RFE]{recursive feature elimination}
\newacro{gp}[GP]{gaussian process}
\newacro{cnn}[CNN]{convolutional neural network}
\newacro{al}[AL]{Apple ladder}
\newacro{sl}[SL]{static ladder}
\newacro{gt}[GT]{ground truth}
\newacro{mlp}[MLP]{multilayer perceptron}
\newacro{vca}[VCA]{video complexity analyzer}
\newacro{uhd}[UHD]{ultra high definition}
\newacro{fhd}[FHD]{full high definition}
\newacro{hd}[HD]{high definition}
\newacro{sd}[SD]{standard definition}
\newacro{plcc}[PLCC]{pearson linear correlation coefficient}
\newacro{srocc}[SROCC]{spearman rank order correlation coefficient}
\newacro{r2}[R2]{R-squared}
\newacro{ssim}[SSIM]{structural similarity index measure}
\newacro{ypsnr}[YPSNR]{Y-peak signal-to-noise ratio}
\newacro{psnr}[PSNR]{peak signal-to-noise ratio}
\newacro{vvenc}[VVenC]{Fraunhofer versatile video encoder}
\newacro{dct}[DCT]{discrete cosine transform}
\newacro{tsn}[TSN]{temporal segment network}
\newacro{abr}[ABR]{adaptive bitrate}
\newacro{e}[E]{spatial energy}
\newacro{h}[h]{temporal energy}
\newacro{iqr}[IQR]{interquartile range}
\newacro{sad}[SAD]{sum of absolute differences}
\newacro{bi-lstm}[Bi-LSTM]{bi-directional long short term memory}
\newacro{ppte}[PPTE]{perceptually-aware online per-title encoding}
\newacro{mse}[MSE]{Mean Squared Error}
\newacro{ra}[RA]{random access}
\newacro{cabac}[CABAC]{context-adaptive binary arithmetic coding}

\newacro{cpu}[CPU]{central processing unit}
\newacro{gpu}[GPU]{graphics processing unit}

\newcommand{\etal}{\emph{et al.}\xspace}

\newcommand{\ie}{\emph{i.e.}, }
\newcommand{\eg}{\emph{e.g.}, }

\newcommand{\xmark}{\ding{55}}

\begin{document}

\title{Convex Hull Prediction Methods for Bitrate Ladder Construction: Design, Evaluation, and Comparison
}

\author{Ahmed Telili, Wassim Hamidouche,~\IEEEmembership{Member,~IEEE}, Hadi Amirpour,~\IEEEmembership{Member,~IEEE}, Sid Ahmed Fezza, Luce Morin,~\IEEEmembership{Member,~IEEE} and Christian Timmerer,~\IEEEmembership{Senior Member,~IEEE}
\thanks{Ahmed Telili, Wassim Hamidouche and Luce Morin are with Univ. Rennes, INSA Rennes, CNRS, IETR--UMR 6164, Rennes, France (e-mail: \href{atelili.at}{\{wassim.hamidouche, ahmed.telili, luce.morin\}@insa-rennes.fr}).}
\thanks{Hadi Amirpour and Christian Timmerer are with Christian Doppler Laboratory ATHENA, Alpen-Adria-Universitat, Klagenfurt, Austria (e-mail: \href{Hadi.Amirpour@aau.at}{\{hadi.amirpour, christian.timmerer\}@aau.at}).}
\thanks{SA. Fezza is with the National Higher School of Telecommunications and ICT, Oran, Algeria (e-mail: \href{sfezza@ensttic.dz}{sfezza@ensttic.dz}).}
\thanks{This work has been supported by Région Bretagne under the DEEPTEC project.}
}
\maketitle

\begin{abstract}
\ac{has} has emerged as a prevalent approach for \ac{ott} video streaming services due to its ability to deliver a seamless user experience. A fundamental component of \ac{has} is the bitrate ladder, which comprises a set of encoding parameters (\eg bitrate-resolution pairs) used to encode the source video into multiple representations. This adaptive bitrate ladder enables the client's video player to
dynamically adjust the quality of the video stream in real-time based on fluctuations in network conditions, ensuring uninterrupted playback by selecting the most suitable representation for the available bandwidth. The most straightforward approach involves using a fixed bitrate ladder for all videos, consisting of pre-determined bitrate-resolution pairs known as \textit{one-size-fits-all}. Conversely, the most reliable technique relies on intensively encoding all resolutions over a wide range of bitrates to build the \textit{convex hull}, thereby optimizing the bitrate ladder by selecting the representations from the convex hull for each specific video. Several techniques have been proposed to predict content-based ladders without performing a costly, exhaustive search encoding. This paper provides a comprehensive review of various convex hull prediction methods, including both conventional and learning-based approaches. Furthermore, we conduct a benchmark study of several handcrafted- and \ac{dl}-based approaches for predicting content-optimized convex hulls across multiple codec settings. The considered methods are evaluated on our proposed large-scale dataset, which includes 300 \acs{uhd} video shots encoded with software and hardware encoders using three state-of-the-art video standards, including \acs{avc}/H.264, \acs{hevc}/H.265, and \acs{vvc}/H.266, at various bitrate points. Our analysis provides valuable insights and establishes baseline performance for future research in this field.
\newline
\textbf{Dataset URL:} \url{https://nasext-vaader.insa-rennes.fr/ietr-vaader/datasets/br_ladder}
\end{abstract}

\begin{IEEEkeywords}
Bitrate ladder, video compression, AVC, HEVC, VVC, rate-quality curves, adaptive video streaming, software/hardware encoding.
\end{IEEEkeywords}

\maketitle

\section{INTRODUCTION}

In recent years, video streaming services, including \ac{vod} and live streaming, have become major contributors to internet traffic. A Sandvine report~\cite{sandvine} reveals that video streaming constituted 65.93\% of total internet traffic in the first half of 2022. Consequently, video service providers have invested significant resources in optimizing video encoding to enhance the \ac{qoe} and ensure seamless streaming performance for all users.

\Ac{has} stands as the state-of-the-art video streaming technology, designed to guarantee the highest possible visual quality delivery at the target bitrate. 
In \ac{has}, the video content is first split into short chunks, called segments, typically ranging from 2 to 10 seconds.  These segments are pre-encoded at various resolutions and quality levels to accommodate a wide range of network conditions, device displays, and computing capabilities. The video segments stored on the server side are transmitted over HTTP to client devices based on their \ac{abr} algorithm~\cite{8424813} requests, which consider their specific bandwidth, display resolution, and computing resource requirements. 
To facilitate the bitrate selection process for each segment, a bitrate ladder is commonly utilized. The bitrate ladder consists of a set of encoding parameters, often referred to as bitrate-resolution pairs, which indicate the encoding configuration for each video segment. These pairs are organized hierarchically, with higher bitrates associated with higher video quality and resolutions. The primary objective of the bitrate ladder is to provide multiple representations of the video, enabling client devices to dynamically select the highest possible quality while minimizing buffering.

\begin{figure}[t!]
\centering
 \includegraphics[width=0.96\linewidth]{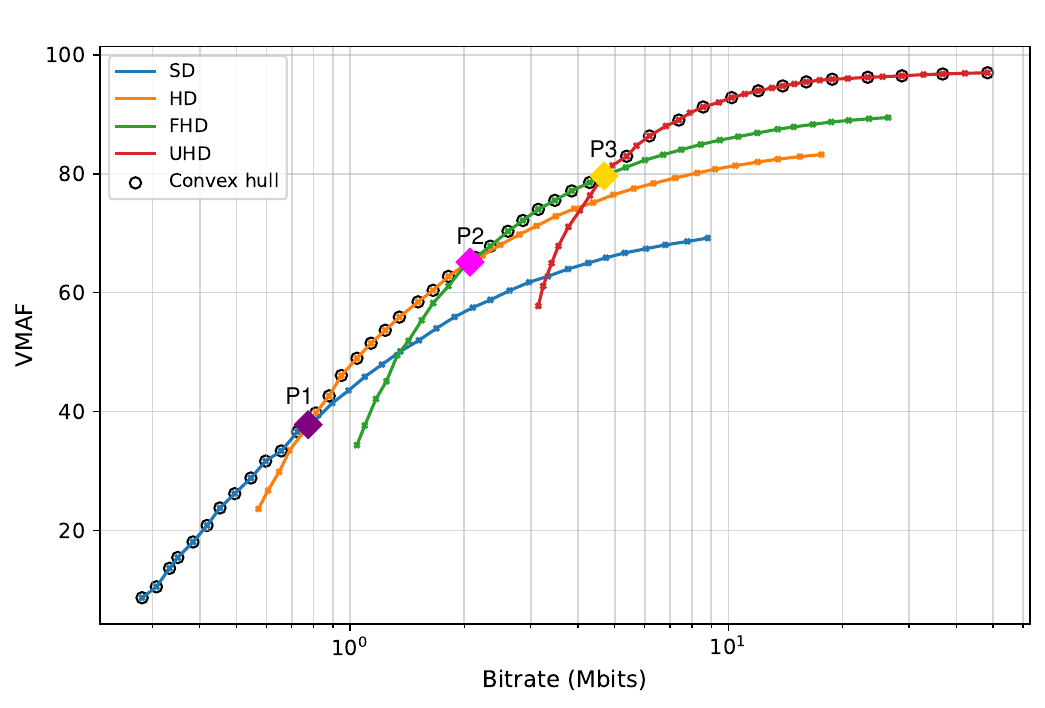}
\caption{Example of rate-distortion curves and their convex hull for the sequence of {\it India-scene-18} from the proposed dataset. P1, P2, and P3 denote the intersection points between SD-HD, HD-FHD, and FHD-UHD resolutions, respectively.}
\vspace{-5mm}
\label{convex_hull}
\end{figure}

One of the classical approaches to constructing a bitrate ladder is to use a static or \textit{one-size-fits-all} ladder, where a fixed set of bitrate-resolution pairs is used for all video content regardless of their characteristics. While easy to implement, this approach often fails to capture the specific characteristics and complexities of individual videos, potentially leading to suboptimal streaming experiences with either wasted bandwidth or degraded quality due to improper bitrate allocation. As a result, more adaptive and content-aware solutions have been proposed. For example, the per-title encoding solution introduced by Netflix~\cite{netflix} is a content-aware approach that constructs a \ac{rd} curve for each individual title (\eg movie, TV show). This is achieved by encoding each title at various bitrates and resolutions, with the resulting quality assessed using an objective full-reference video quality metric. Fig.~\ref{convex_hull} illustrates \ac{rd} curves for a single video title, where distortion is measured using the \ac{vmaf} metric~\cite{noauthor_vmaf_nodate}. Subsequently, the convex hull approach was refined with per-shot encoding \cite{netflix2}, where each video is segmented into shots with similar frame characteristics that respond similarly to varying encoding parameters. This allows for the creation of customized bitrate ladders for each shot, further optimizing the adaptive convex hull approach. However, generating convex hulls for adaptive streaming, whether per video shot or title, remains computationally expensive due to the vast parameter space encompassing resolution, quality level, codec type, and more. This complexity translates into a time-consuming and resource-intensive process, rendering it costly for \ac{vod} streaming and impractical for live streaming scenarios.

To address these challenges, numerous \ac{ml}-based convex hull prediction techniques have been proposed by both academic and industry experts in recent years \cite{goswami2018adaptive,bhat2020case,8456297,8954529,katsenou2021vmaf,9749811,nasiri2022ensemble,menon2022perceptually,huang2021deep,bitmovin,8361841,10668820,9173554,10422719,10222094}. These techniques aim to predict per-scene or per-shot convex hulls, eliminating the need for exhaustive encoding. In this context, we have conducted a benchmark study that compares various handcrafted- and \ac{dl}-based methods, providing an in-depth analysis of their complexity and prediction performance. In addition, our benchmark incorporates a multi-codec approach to enhance the evaluation, allowing for a more comprehensive assessment of methods under different encoding scenarios. This comparison and analysis will contribute to a better understanding of the strengths and weaknesses of handcrafted- and \ac{dl}-based techniques, ultimately guiding the development and comparison of more efficient and adaptable video streaming solutions. The principal contributions of this paper are listed as follows:

\begin{itemize}
  \item A systematic classification and extensive review of current bitrate ladder construction methods for adaptive video streaming, including a detailed analysis of conventional and learning-based techniques.
  \item A large dataset with 300 Ultra-HD video sequences, encoded through hardware and software video encoders using three standards: \acs{avc}/H.264~\cite{wiegand2003overview}, \acs{hevc}/H.265~\cite{sullivan2012overview}, and \acs{vvc}/H.266~\cite{bross2021overview} in four resolutions and multiple \ac{qp} values.
  \item Exploring several handcrafted and \ac{dl}-based features for both \ac{vod} and live streaming scenarios with various \ac{ml} models for convex hull prediction.
  \item A comprehensive benchmark study to assess the performance of \ac{ml}-based methods, along with an in-depth analysis of prediction performance and complexity overhead on \acs{cpu} and \acs{gpu} platforms.
\end{itemize}

The remainder of this paper is organized as follows. Section~\ref{sec:review} gives a comprehensive review of bitrate ladder prediction methods. Furthermore, the proposed dataset is presented and characterized in Section~\ref{sec:review}. Section~\ref{benchmark} presents the development of benchmarking evaluation, while Section~\ref{results} provides and analyses the experimental results. Finally, Section~\ref{conclusion} concludes this paper.

\section{BITRATE LADDER CONSTRUCTION METHODS}
\label{sec:review}
This section presents a systematic classification and comprehensive review of existing methods for constructing bitrate ladders in adaptive video streaming. We categorize these methods into two primary classes: (\textit{i}) static methods, which disregard content characteristics, and (\textit{ii}) dynamic methods, which involve adaptive selection of bitrate ladders based on content characteristics, network conditions, and other relevant factors. To circumvent computationally intensive brute-force approaches for finding optimal content-aware bitrate ladders, recent dynamic methods typically employ \ac{ml} techniques to predict the optimized convex hull based on extracted content features. These \ac{ml}-based approaches are further classified into two types: (\textit{a}) methods utilizing handcrafted feature extraction and (\textit{b}) methods employing deep feature extraction relying on deep neural networks.

\begin{figure}[t!]
\centering
 \includegraphics[width=0.98\linewidth]{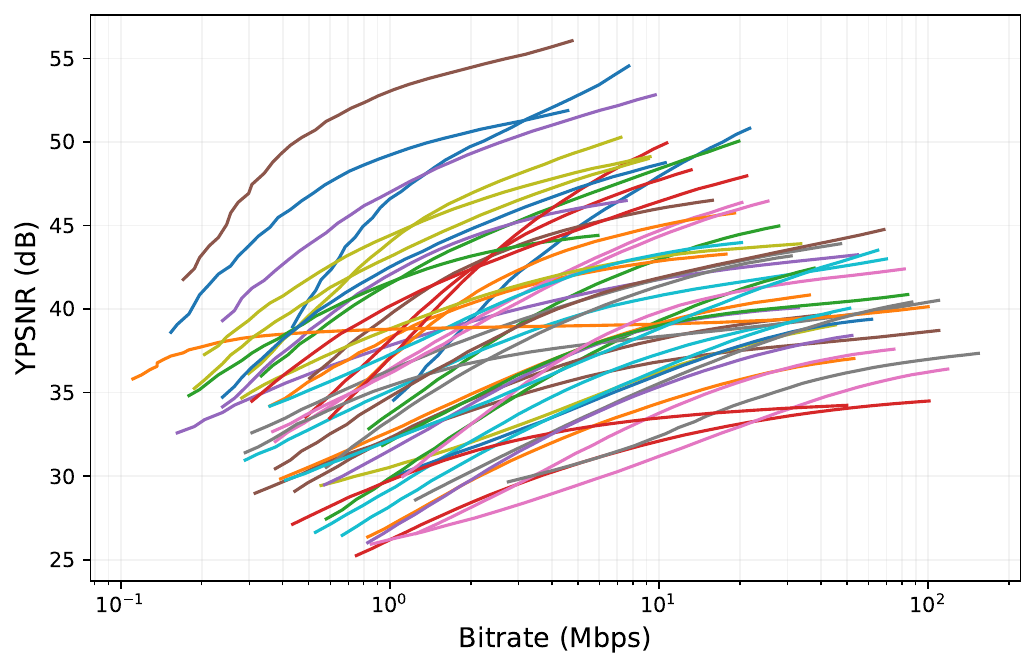}
\caption{\Ac{rd} curves for 50 randomly sampled footage at 1080p resolution.}
\vspace{-5mm}
\label{static_fhd}
\end{figure}

The static (\ie content-independent) bitrate ladder, also known as the \textit{one-size-fits-all} bitrate ladder, is the conventional approach that offers predefined recommendations for encoding parameters, such as bitrate and resolution, for a given video content. Apple proposed one commonly adopted static bitrate ladder in Tech Note TN2224~\cite{apple_tech}. Similarly, other video streaming platforms, such as YouTube~\cite{Google} and Twitch~\cite{Twitch}, have also provided their recommendations on encoding settings for their streamers. 
Although these \textit{one-size-fits-all} bitrate ladders are simple to implement and straightforward to use, they have limitations in providing optimal encoding parameters for different types of videos.  For complex videos, such as those with high-motion scenes, these static ladders may allocate insufficient bitrate, resulting in significant blocking and other visual artifacts. 
On the other hand, for less complex video content, such as cartoons, these ladders might over-allocate bitrate, leading to storage and bandwidth wastage. For instance, in Fig.~\ref{static_fhd}, we illustrate the \ac{rd} curves of 50 video sequences at 1080p resolution, encoded using the x265 software video encoder across a range of q\ac{qp} values. The figure demonstrates the substantial variability in compression efficiency across different video sequences. Some sequences achieve high \ac{ypsnr} of 45 dB or more at bitrates as low as 1 Mbps, while others require bitrates exceeding 20 Mbps to reach a satisfactory \ac{ypsnr} of 38 dB. This inherent diversity in video content underscores the limitations of static bitrate ladder schemes, which cannot optimally cater to the unique characteristics of individual titles. Consequently, more sophisticated dynamic techniques have been developed to address this challenge.

\subsection{Static methods}
\label{sec:static}

\subsection{Dynamic methods}
In dynamic methods, all parameters of the bitrate ladder, including the number of representations and their corresponding bitrate, resolution, and other encoding parameters, are determined dynamically based on both content characteristics and prevailing network conditions. These methods can be broadly categorized into conventional methods (\ie do not involve any learning process), and learning-based methods, which leverage machine learning to optimize bitrate selection.
\subsubsection{Conventional methods}
Conventional methods typically rely on intensive brute-force or rule-based techniques. For instance, 
Netflix pioneered the per-title approach~\cite{netflix}, encoding videos at various quality levels and resolutions to construct the convex hull of \ac{rd} curves using the \ac{vmaf} metric. This customized bitrate ladder, tailored to each video's unique content, improves the \ac{qoe}~\cite{zhang_deepqoe_2020} compared to static approaches. As the Netflix service expanded, an optimized per-chunk cloud-based encoding method was introduced~\cite{7532605} for enhanced scalability and reliability. This method identifies suitable bitrate-resolution pairs for each video chunk by evaluating their visual complexity through a \ac{crf}-based multi-pass encoding process. However, both methods are suboptimal as titles and chunks can contain scenes with varying visual complexity. To address this limitation, Netflix further refined their approach~\cite{netflix2} by dividing videos into shots, which are groups of frames with similar encoding behavior. Individual processing of each shot, using the original exhaustive encoding approach, enables the creation of per-shot bitrate ladders. Similarly, Chen~\etal~\cite{chen2018optimized} extracted \ac{rd} characteristics through multiple encodes of each video chunk at various resolutions and bitrates. Unlike Netflix, this approach considers not only content characteristics but also user bandwidth and viewport size distributions when selecting bitrate-resolution pairs.

Reznik~\etal~\cite{reznik2018optimal} introduced an analytical approach that incorporates both content behavior and network statistics to construct the bitrate ladder, formulating it as a nonlinear optimization problem. They utilized the \ac{ssim}~\cite{wang_image} as the quality metric and employed probabilistic models to set bitrate constraints. This approach was later extended to accommodate multiple codec types, including \acs{avc}/H.264 and \acs{hevc}/H.265~\cite{reznik2019optimal}.

Toni~\etal~\cite{toni_optimal_2015} proposed an optimization framework for selecting adaptive video streaming representations to enhance user satisfaction under network constraints. Their integer linear programming model maximizes user satisfaction by considering video content, network capacity, and user characteristics, outperforming traditional vendor recommendations in fairness and outage reduction. Tashtarian~\etal~\cite{tashtarian2023lalisa} developed LALISA, an online lightweight framework for optimizing bitrate ladders in HTTP-based live video streaming. LALISA dynamically determines the optimal bitrate ladder to minimize encoding and delivery costs while maintaining satisfactory \ac{qoe} during live sessions. 
The work has been extended in~\cite{tashtarian_artemis_2024} by introducing a mega bitrate ladder to clients, gathering their desired bitrate requests, and encoding only a subset of the mega bitrate ladder—based on the probability of requests—as the main bitrate ladder. This approach resolves the issue faced by LALISA, where modifications to the clients' \ac{abr} algorithms were required.  Furthermore, Lebreton~\etal~\cite{lebreton_quitting_2023}~proposed a bitrate ladder estimation method based on the likelihood of user quitting, incorporating historical network throughput data to define an ABR-control agnostic bitrate ladder estimation method. 
Additionally, Amirpour~\etal~\cite{amirpour2021pstr} proposed the \ac{pstr} method, incorporating temporal resolution (specifically, framerate) as an additional dimension in bitrate ladder construction. This technique involves encoding each video chunk at multiple resolutions and framerates to identify the optimal settings for each bitrate. While introducing temporal resolution into per-title encoding increases complexity, considering both spatial and temporal resolutions can yield substantial bitrate savings.
Further advancing this work, Amirpour~\etal~\cite{amirpour2022deepstream} presented DeepStream, a scalable content-aware per-title encoding system supporting both \acs{cpu}-only and \acs{gpu}-equipped users. DeepStream comprises two layers: a base layer that constructs a bitrate ladder using existing per-title encoding methods and an enhancement layer that improves the quality through content-aware deep video super-resolution. This enhancement leverages lightweight \ac{cabac} for efficient \ac{dnn} compression, ensuring backward compatibility for \acs{cpu}-only users while providing enhanced perceived video quality for those with \acs{gpu} capabilities. Additionally, Ghasempour~\etal~\cite{ghasempour_energy-aware_2024} introduced energy as a new parameter for resolution selection, incorporating decoding energy consumption alongside quality considerations. They further extended their work~\cite{ghasempour_energy-aware_2024} by including frame rate as an additional factor, complementing the spatial resolution. Toni~\etal~\cite{toni_optimal_2017} proposed a representation set optimization problem for multiview adaptive streaming systems.

\subsubsection{Learning-based methods}
Learning-based methods constitute a class of dynamic techniques that leverage machine learning to identify patterns within input video data, enabling the construction of content-dependent bitrate ladders, specifically convex hulls, without the computational burden of brute-force encoding across all representations. These methods can be broadly categorized into two distinct types based on their feature extraction approach: those employing handcrafted features and those utilizing deep learning for feature extraction.

\vspace{2mm}
\textbf{Handcrafted features. }
In learning-based methods, the first class focuses on approaches that leverage handcrafted features. These features are meticulously designed and manually derived from the video content to model.  
For instance, Bhat \textit{et al.}~\cite{bhat2020case} introduced an innovative approach to swiftly predict adaptive spatial resolutions without requiring multiple encodings. Their algorithm incorporates an extensive array of features, covering rate control-based factors (\ac{qp}, frame rate, targeted bitrate), spatial features (histogram of pixel values, variance of \acl{ctu}), temporal features (motion vectors, scene change scores), and encoder pre-analysis-based features (estimated \ac{qp} of all frames in the initial \ac{gop}, estimated intra probability). From the set of features, 43 input features are derived and fed into a machine learning classifier, \ac{mlp} or \ac{rf}, for resolution prediction.

In~\cite{8456297}, the authors introduced a perceptual quality-driven method for generating adaptive encoding ladders. This method employs the \ac{jnd} scale, representing the smallest perceptible change in visual quality for human viewers. Video sequences are pre-encoded at various resolutions and three \acp{qp} per resolution, and features like bitrates, \ac{psnr}, \ac{ssim}, and \ac{vmaf} are extracted. \Ac{svr} is then used to estimate \ac{jnd} scores based on these features, enabling optimization of the encoding ladder to minimize the number of perceptually similar bitrate-resolution pairs.

Menon~\etal~\cite{menon2022perceptually} adopted a similar approach in their \ac{ppte} scheme for live streaming. They predict optimal bitrate-resolution pairs for each video segment based on \ac{jnd}, using low-complexity \ac{dct}-energy-based features to assess spatial and temporal complexity. This enables the prediction of bitrates where \acp{jnd} occur. Notably, unlike~\cite{8456297}, this approach does not rely on pre-encoding, thus accelerating the bitrate ladder construction process.

Katsenou~\etal~\cite{8954529} introduced a content-agnostic approach using \ac{gpr} to predict cross-over \acp{qp} for different spatial resolutions in \ac{hevc} encoding. The authors employed handcrafted spatio-temporal features, such as \ac{glcm} and \ac{tc}, extracted from uncompressed \ac{uhd} video sequences to predict \acp{qp} based on \ac{psnr}. In a subsequent work, Katsenou~\etal~\cite{katsenou2021vmaf} expanded their approach to incorporate the \ac{vmaf} metric as an additional quality criterion. Similarly, Silhavy~\etal~\cite{9749811}  utilized machine learning algorithms, including \ac{svr}, \ac{mlp}, and \ac{rfr}, to construct a content-aware bitrate ladder using \ac{vmaf} as the quality metric. Another handcrafted feature-based method was presented in~\cite{nasiri2022ensemble}, employing two supervised machine learning algorithms, regression, and a classification method—trained on spatio-temporal features (\acs{glcm}, \acs{tc}) extracted from uncompressed video sequences. An ensemble aggregation technique was applied to predict a bitrate ladder for the \ac{vvc} encoder, enhancing the performance of both \ac{ml} algorithms. Additionally, Menon~\etal~\cite{menon_opte_2022} introduced an online resolution prediction approach that leverages low-complexity \ac{dct}-energy-based spatio-temporal features extracted from video sequences to predict the optimal resolution for a target bitrate based on these features. Similarly, Qin~\etal~\cite{qin_content_2023} utilize \ac{si} and \ac{ti} to extract spatio-temporal features, which are then fed into a \ac{svm} to determine whether to encode a video at its original resolution or downsample it. The work has been extended in~\cite{yang_content_2024} to incorporate both resolution and frame rate downsampling. Furthermore, Adhuran and Kulupana~\cite{adhuran_content-aware_2023} proposed a two-stage machine learning framework for content-aware convex hull generation in video compression, which uses spatio-temporal features such as motion vectors, GLCM metrics, and no-reference quality measures. Their framework employs Gradient Boost Regressors and Random Forest Regressors to predict compressed domain features and bitrates, enabling efficient bitrate ladder construction and significant compression improvements.

\vspace{2mm}

\textbf{Deep features. }
In contrast to methods reliant on handcrafted features, the deep features category encompasses techniques that leverage \acp{dnn} for extracting relevant features from video content. For example, Huang~\etal~\cite{huang2021deep} proposed DeepLadder, a reinforcement learning-based method that considers video content, network traffic capacity, and storage costs to determine suitable encoder settings for each resolution. Notably, DeepLadder utilizes a pre-trained ResNet-50 model as its backbone for spatial feature extraction from intra-frames.

Furthermore, Xing~\etal~\cite{xing2019predicting} introduced a content-adaptive rate control solution employing a \ac{tsn}~\cite{wang2016temporal} to extract spatio-temporal features from video content. These features are then processed by a fully connected layer followed by a Softmax function to predict the optimal rate-control target based on content characteristics. This deep learning model accommodates both average bitrate control and \ac{crf} modes, predicting the optimal bitrate in the former and the rate factor in the latter. 

Alternatively, a \ac{dnn}-based method presented in~\cite{paul_convex_2024} employs a \ac{rcn} to estimate the convex hull of video shots. This method effectively captures long-term spatio-temporal dependencies across video shots by jointly processing spatial and temporal information using \ac{crgu}~\cite{ballas2015delving}. Notably, the prediction task is formulated as a multi-label classification problem rather than a traditional regression problem. Additionally, a transfer learning technique is employed to mitigate the limited availability of uncompressed public video datasets suitable for training deep models.

Given the data-intensive nature of many learning-based solutions, we dedicate the subsequent section to the construction of a large-scale dataset specifically tailored to support our research objectives.

\section{DATASET CONSTRUCTION}

\begin{figure*}[!]
\centering
\scriptsize
\centering
\begin{minipage}[b]{0.33\linewidth}
\centering
\centerline{\includegraphics[width=1\linewidth]{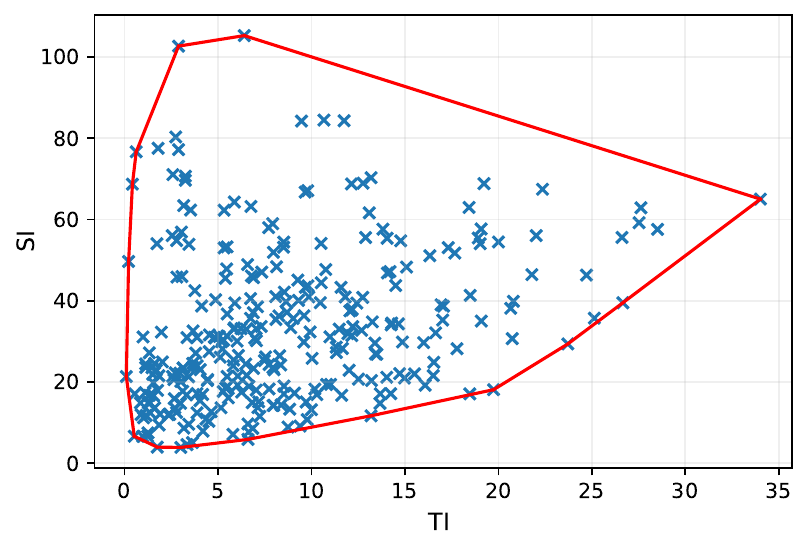}}
(a) SI vs. TI
\end{minipage}
\begin{minipage}[b]{0.33\linewidth}
\centering
\centerline{\includegraphics[width=1\linewidth]{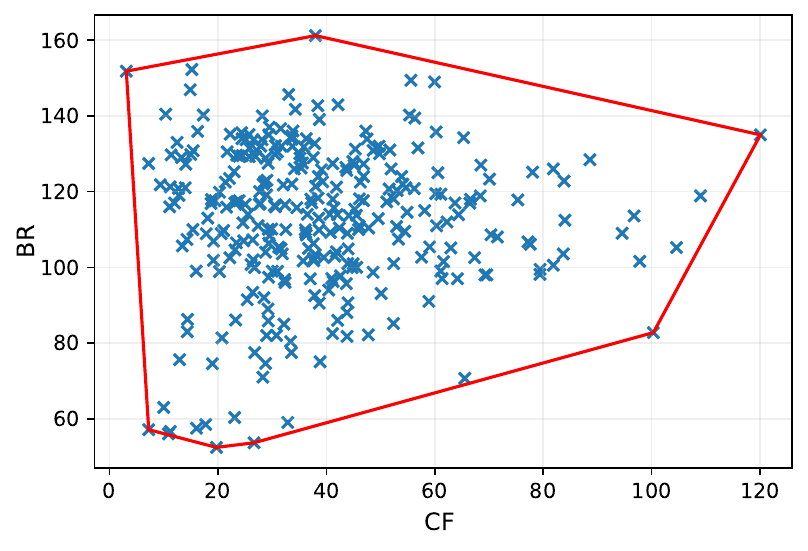}}
(b) BR vs. CF
\end{minipage}
\begin{minipage}[b]{0.33\linewidth}
\centering
\centerline{\includegraphics[width=1\linewidth]{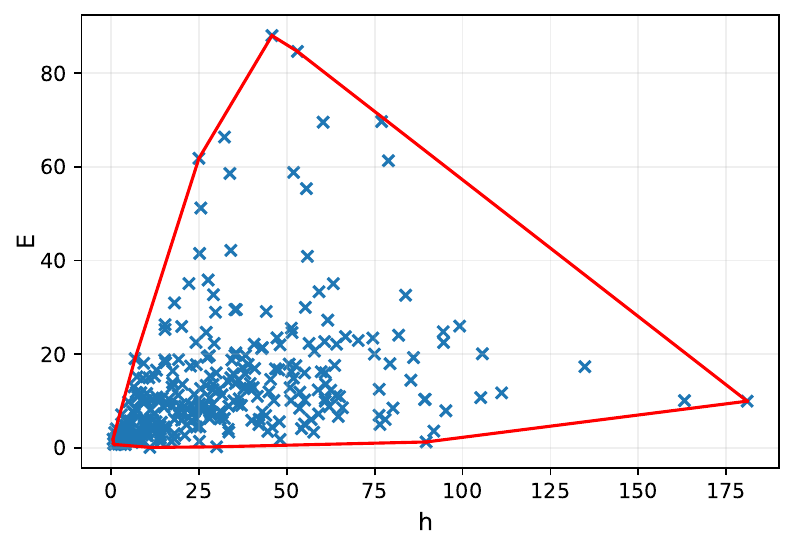}}
(b) h vs. E
\end{minipage}
\caption{Source content distribution in paired feature space with corresponding convex hulls. Left column: TI×SI, middle column: CF×BR and
right column: hxE.}
\label{features_plots}
\end{figure*}

\subsection{Convex hull construction}
\label{convex_construction}

A diverse and extensive video dataset is crucial for any data-driven approach. Thus, in this section, we present the \ac{avsd}, a comprehensive and diversified collection of 300 UHD video shots from publicly available sources. Additionally, we provide the computed convex hulls for these shots, following encoding using three distinct video coding standards: \ac{avc}/H.264, \ac{hevc}/H.265, and \ac{vvc}/H.266. We firmly believe that the availability of the \ac{avsd} dataset will be of immense value and contribute to the advancement of research in adaptive video streaming and related domains.

\subsection{Video collection}
Researchers in the field of learned video compression often face a challenge due to the limited availability of publicly accessible, high-quality, uncompressed video content. To mitigate this issue, we have meticulously curated a dataset of 300 videos, including both uncompressed video sequences and high-quality compressed versions. Within this collection, we have assembled 120 uncompressed video sequences from a variety of sources, including Netflix Chimera~\cite{ioannis2015Netflix}, AWS Elemental~\cite{AWSElemental}, MCML~\cite{cheon2017subjective}, Harmonic~\cite{harmonicFootage}, Ultra Video Group~\cite{tampereTest}, SJTU~\cite{song2013sjtu}, and the Waterloo IVC 4K Video Quality Database~\cite{li2019avc}. The remaining 180 sequences were thoughtfully selected from the YouTube-UGC dataset~\cite{wang2019youtube}, exclusively featuring high-quality content. While many of these sequences initially boasted a resolution of 4096$\times$2160, they were subsequently cropped to 3840$\times$2160 and converted into a 4:2:0 chroma subsampled format, if not originally formatted as such. To ensure homogeneity, all sequences underwent a scene segmentation process, guaranteeing that each sequence comprises only a single scene. As a final step, all sequences were temporally cropped to 64 frames, resulting in a dataset that offers a comprehensive view of different scenes and scenarios.

\subsection{Dataset characterization}
Winkler {\it et al.}~\cite{6280595} originally proposed three video descriptors: spatial activity, temporal activity, and colorfulness, to characterize content diversity within datasets. In our study, we extend this analysis of content diversity by incorporating six low-level features. These features encompass \ac{si}~\cite{itu_t_p910}, \ac{ti}~\cite{itu_t_p910}, \ac{br}~\cite{fairchild2013color}, \ac{cf}~\cite{fairchild2013color}, and two novel features derived from \ac{vca}~\cite{menon2022vca}: spatial complexity (E) and temporal complexity (h). Each of these features is computed individually for every frame in the dataset. Subsequently, the computed values are averaged to obtain an overall (mean) score. To visualize the feature coverage of our dataset, we present scatter plots with convex hulls for paired features, as depicted in Fig.~\ref{features_plots}. The video sequences in our dataset span a wide range of the spatio-temporal domain, with \ac{si} values ranging from 5 to 105 and \ac{ti} values from 0 to 35. This diversity extends to the spatio-temporal complexities found within the dataset. The majority of videos exhibit low complexity, characterized by single-shot scenes. However, a smaller fraction contains highly complex shots, adding variety to the dataset. Additionally, the scatter plot of \ac{br} versus \ac{cf} unveils a rich spectrum of content types. \Ac{br} values span from 60 to 160, indicating diverse lighting conditions and scene settings. Further, \ac{cf} values range from 0 to 120, signifying a multitude of color palettes and visual styles.

To determine the \ac{rd} of video sequences and construct the ground truth bitrate ladder, we downscaled all sequences to \ac{fhd}, \ac{hd}, and \ac{sd} using a Lanczos-3 filter~\cite{duchon1979lanczos}. Subsequently, we encoded the video sequences in four resolutions, which included both the downsampled versions and the original \ac{uhd} sequences, denoted as $\mathcal{S} \in \{720~\times~480, 1280~\times~720, 1920~\times~1080, 3840~\times~2160\}$. All videos were encoded at 60 fps. These resolutions represent typical choices in streaming applications~\cite{dacast}. The encoding process was performed using three video coding standards: \ac{avc}/H.264, \ac{hevc}/H.265, and \ac{vvc}/H.266. For \ac{avc}/H.264 and \ac{hevc}/H.265 standards,  we employed both software (x264/x265) and hardware-based (NVENC) implementations in \ac{ra} configuration, using a range of constant \ac{qp} values from $\mathcal{Q} \in \{15, 16, \dots, 44, 45\}$ at medium preset. The encoding process for these codecs was conducted using FFmpeg version 5.0~\cite{ffmpeg}. However, since no public hardware encoder for \ac{vvc}/H.266 is currently available,  we exclusively utilized its software encoder,~\ac{vvenc} version 1.6~\cite{wieckowski2021vvenc}. Similar to the other codecs, we operated the software encoder in random access configuration with a set of constant \ac{qp} values from $\mathcal{Q}\prime \in \{16, 18, \dots, 46, 48\}$. This process generated 564 bitstreams per source content, totaling 169,200 encoded video shots. Further, it is essential to mention that the software encoding process was performed on a 16-core, 3.70 GHz Intel Xeon W-2145 CPU with 64 GB RAM, while hardware encoding was performed on an NVIDIA GeForce RTX 2080 Ti \acs{gpu}. Following this, all the encoded bitstreams underwent a decoding process and were upsampled back to their original resolution (2160p) using the same filter. Subsequently, we proceeded to construct the \ac{rd} curves. To do this, we measured the objective video quality of the decoded sequences using two full-reference quality metrics: Y-\ac{psnr} and \ac{vmaf}. Finally, we defined the intersection points between the  \ac{rd} curves of the same video across different resolutions to create the convex hull. Specifically, we identified these points as \ac{sd}-\ac{hd}, \ac{hd}-\ac{fhd}, and \ac{fhd}-\ac{uhd}, denoted as P1, P2, and P3, respectively. These points are visually represented in Fig.~\ref{convex_hull} for a specific video shot encoded with the x264 software encoder. In this study, we address scenarios where direct intersections between \ac{rd} curves of consecutive resolutions may not occur. We hypothesize that if a resolution \( \mathcal{S}_{i+1} \) consistently delivers higher quality than resolution \( \mathcal{S}_i \), the intersection point is defined as the starting bitrate of \( \mathcal{S}_{i+1} \), corresponding to the lowest \ac{qp}. Conversely, if \( \mathcal{S}_i \) remains superior across the bitrate spectrum, we designate the intersection point at a maximum predefined bitrate value.


\begin{table}[t]
\caption{List of VoD-HandC features and their statistics.}
\centering
\adjustbox{max width=0.48\textwidth}{%
\label{features_table}
\begin{tabular}{@{}
>{\columncolor[HTML]{FFFFFF}}l
>{\columncolor[HTML]{FFFFFF}}l @{}}
\toprule 
{\color[HTML]{1D1D1F} Features}                                                                        & {\color[HTML]{1D1D1F} Statistics}                                                                                                                                                                         \\ \midrule
{\color[HTML]{1D1D1F} \begin{tabular}[c]{@{}l@{}}Grey-Level Co-occurrence Matrix\\ (GLCM)\end{tabular}} & {\color[HTML]{1D1D1F} \begin{tabular}[c]{@{}l@{}}F1.$\textrm{meanGLCM}_{\textrm{cor}}$, F2.$\textrm{stdGLCM}_{\textrm{cor}}$,  \\ F3.$\textrm{meanGLCM}_{\textrm{con}}$, F4$\textrm{stdGLCM}_{\textrm{con}}$,  \\ F5.$\textrm{meanGLCM}_{\textrm{enr}}$, F6.$\textrm{stdGLCM}_{\textrm{enr}}$, \\ F7.$\textrm{meanGLCM}_{\textrm{hom}}$, F8.$\textrm{stdGLCM}_{\textrm{hom}}$, \\ F9.$\textrm{meanGLCM}_{\textrm{ent}}$, F10.$\textrm{stdGLCM}_{\textrm{ent}}$\end{tabular}} \\ \midrule
{\color[HTML]{1D1D1F} \begin{tabular}[c]{@{}l@{}}Temporal Coherence\\ (TC)\end{tabular}}               & {\color[HTML]{1D1D1F} \begin{tabular}[c]{@{}l@{}}F11.$\textrm{meanTC}_{\textrm{mean}}$, F12.$\textrm{meanTC}_{\textrm{std}}$, \\ F13.$\textrm{stdTC}_{\textrm{mean}}$, F14.$\textrm{stdTC}_{\textrm{std}}$, \\ F15.$\textrm{meanTC}_{\textrm{skw}}$, F16.$\textrm{stdTC}_{\textrm{skw}}$, \\ F17.$\textrm{meanTC}_{\textrm{kur}}$, F18.$\textrm{stdTC}_{\textrm{kur}}$, \\ F19.$\textrm{meanTC}_{\textrm{entr}}$, F20.$\textrm{stdTC}_{\textrm{entr}}$\end{tabular}}              \\ \midrule
{\color[HTML]{1D1D1F} Spatial Information (SI)}                                                        & {\color[HTML]{1D1D1F} F21.mean$_{SI}$, F22.std$_{SI}$}                                                                                                                                                              \\ \midrule
{\color[HTML]{1D1D1F} \begin{tabular}[c]{@{}l@{}}Temporal Information  (TI)\end{tabular}}             & {\color[HTML]{1D1D1F} F23.mean$_{TI}$, F24.std$_{TI}$}                                                                                                                                                              \\ \midrule
{\color[HTML]{1D1D1F} Colorfulness (CF)}                                                               & {\color[HTML]{1D1D1F} F25.mean$_{CF}$, F26.std$_{CF}$}                                                                                                                                                              \\ \midrule
{\color[HTML]{1D1D1F} Noise estimation}                                                                           & {\color[HTML]{1D1D1F} F27.mean$_{Noise}$, F28.std$_{Noise}$}                                                                                                                                                       \\ \midrule
{\color[HTML]{1D1D1F} Normalized Cross Correlation (NCC)}                                              & {\color[HTML]{1D1D1F} F29.mean$_{NCC}$, F30.std$_{NCC}$}                                                                                                                                                            \\ \bottomrule 
\end{tabular}
}\vspace{-3mm}

\end{table}
\section{CONVEX HULL PREDICTION TECHNIQUES}
\label{benchmark}

This section describes our benchmark for evaluating learning-based methods for convex hull prediction in adaptive video streaming. The selection of bitrate-resolution pairs from the convex hull is not addressed in this work. We focus on two categories: (\textit{i}) methods based on handcrafted features and (\textit{ii}) those based on deep neural networks, aiming to provide valuable insights into their strengths and limitations.  These insights aim to guide researchers and practitioners toward more efficient solutions.

\subsection{Methods based on handcrafted features}
In this subsection, we examine two sets of handcrafted features for predicting convex hulls. The first set includes classic low-level features successfully used in compression and video streaming-related research~\cite{8954529, afonso2018spatial, 10018038}. These classic features, although widely used, are computationally intensive and time-consuming to extract, particularly for \ac{uhd} videos. Therefore, they are more suitable for \ac{vod} streaming applications, hence referred to as VoD-HandC features. On the other hand, the \ac{vca} features, referred to as Live-HandC features,  aim to provide a more efficient alternative for live-streaming scenarios with real-time and low latency constraints. The results of these methods, including their complexity, will be further explored in Section~\ref{results}.

\subsubsection{VoD-HandC features} The VoD-HandC feature extraction process can be formulated as follows:
\begin{equation}
\mathcal{S}_{F_i} = \Psi_{\text{VoD}} (\mathcal{V}),
\end{equation}
where $\mathcal{S}_{F_i}$ represents the extracted features set, the $\Psi_{\text{VoD}}$ represents VoD-HandC feature extraction function, and  $\mathcal{V}$ denotes the input video sequence.  \\ 
We have selected a set of spatio-temporal features from various low-level features successfully used in our previous related work \cite{10018038}. Table~\ref{features_table} provides the full set of these features and their statistics.

\begin{table}[]
\centering
\caption{List of Live-HandC features and their statistics.}
\label{VCA_feat}
\resizebox{\columnwidth}{!}{%
\begin{tabular}{@{}lll@{}}
\cmidrule(r){1-2}
Features                                                  & Statistics                                                                                     &  \\ \cmidrule(r){1-2}
\multirow{4}{*}{Spatial energy (E)}                       & F1.mean$_{E}$, F2.std$_{E}$, F3.min$_{E}$,                                &  \\
                                                          & F4.max$_{E}$, F5.$25^{th}$$_{E}$, F6.$50^{th}$$_{E}$,                               &  \\
                                                          & F7.$75^{th}$$_{E}$, F8.iqr$_{E}$, F9.skw$_{E}$,                                &  \\
                                                          & F10.kur$_{E}$                                                                           &  \\ \cmidrule(r){1-2}
\multirow{4}{*}{Temporal energy (h)}                      & F11.mean$_{h}$, F12.std$_{h}$, F13.min$_{h}$,                             &  \\
                                                          & F14.max$_{h}$, F15.$25^{th}$$_{h}$, F16.$50^{th}$$_{h}$,                            &  \\
                                                          & F17.$75^{th}$$_{h}$, F18.iqr$_{h}$, F19.skw$_{h}$,                             &  \\
                                                          & F20.kur$_{h}$                                                                           &  \\ \cmidrule(r){1-2}
\multirow{4}{*}{Gradient of temporal energy ($\epsilon$)} & F21.mean$_{\epsilon}$, F22.std$_{\epsilon}$, F23.min$_{\epsilon}$,  &  \\
                                                          & F24.max$_{\epsilon}$, F25.$25^{th}$$_{\epsilon}$, F26.$50^{th}$$_{\epsilon}$, &  \\
                                                          & F27.$75^{th}$$_{\epsilon}$, F28.iqr$_{\epsilon}$, F29.skw$_{\epsilon}$  &  \\
                                                          & F30.kur$_{\epsilon}$                                                                  &  \\ \cmidrule(r){1-2}
\multirow{4}{*}{Brightness (BR)}                          & F31.mean$_{BR}$, F32.std$_{BR}$, F33.min$_{BR}$,                          &  \\
                                                          & F34.max$_{BR}$, F35.$25^{th}$$_{BR}$, F36.$50^{th}$$_{BR}$,                         &  \\
                                                          & F37.$75^{th}$$_{BR}$, F38.iqr$_{BR}$, F39.skw$_{BR}$,                          &  \\
                                                          & F40.kur$_{BR}$                                                                          &  \\ \cmidrule(r){1-2}
\end{tabular}
}\vspace{-5mm}
\end{table}

\vspace{2mm}
\noindent {\bf \Ac{glcm}}~\cite{haralick1973textural} is a spatial feature that analyzes the intensity contrast between adjacent pixels to characterize the texture and patterns in an image. \ac{glcm} comprises five fundamental descriptors: contrast, correlation, homogeneity, energy, and entropy. \vspace{2mm}
\newline
{\bf \Ac{tc}}~\cite{katsenou2016predicting} is a temporal feature employed to characterize the predictability of a frame based on its predecessor, assessing the consistency of spectral amplitude between consecutive images to reflect temporal prediction complexity. In practical terms, \ac{tc} is computed for each pair of successive frames.\vspace{2mm}
\newline
{\bf \Ac{si}}~\cite{itu_t_p910} is a feature employed to gauge the level of spatial detail within an image, encompassing textures, patterns, and structures. \ac{si} quantifies the associations among pixel intensity values within an image, offering insights into local variations in the image. \vspace{2mm}
\newline
{\bf \Ac{ti}}~\cite{itu_t_p910} is a feature used to evaluate the video's temporal characteristics, such as motion, scene changes, and object dynamics. By quantifying the relationships between consecutive frames, \ac{ti} reveals changes over time.   \vspace{2mm}
\newline
{\bf Colorfulness (CF)}~\cite{fairchild2013color} is a spatial feature employed to evaluate the richness and diversity of colors within a picture. It quantifies the degree of color variation and saturation, offering insights into the visual appeal and vibrancy of the content. \vspace{2mm}
\newline
{\bf Noise estimation}~\cite{liu2006noise} is a spatial feature employed to evaluate the presence and intensity of noise within an image. It measures the extent of random variations, artifacts, or distortions in pixel values, providing insight into the overall quality and sharpness of the content.   \vspace{2mm}
\newline
{\bf \Ac{ncc}}~\cite{lewis1995fast}  is a similarity measure used to assess the temporal similarity between frames based on the correlation of pixel intensity values between two successive frames. To analyze \ac{ncc} across an entire sequence, the \ac{ncc} is computed for each pair of consecutive frames. 

For each feature, sequence-level statistics are computed to assess the overall characteristics of the video sequence, as indicated in Table~\ref{features_table}.

\begin{table*}[t]
\centering
\caption{Selected features of predicted cross-over bitrates for RD curves based on YPSNR.}
\label{YPSNR_selec}
\resizebox{\textwidth}{!}{%
\begin{tabular}{@{}lllllll@{}}
\toprule
\multirow{2}{*}{\begin{tabular}[c]{@{}l@{}}Cross-over\\ bitrates\end{tabular}} & Codec                                                      & \multicolumn{2}{c}{AVC}                                                                                                                                                                           & \multicolumn{2}{c}{HEVC}                                                                                                                                                                                & \multicolumn{1}{c}{VVC}                                                                                 \\ \cmidrule(l){2-2} \cmidrule(l){3-4} \cmidrule(l){5-6} \cmidrule(l){7-7}
                                                                               & Plateform                                                  & \multicolumn{1}{c}{Software (FFmpeg)}                                                                       & \multicolumn{1}{c}{Hardware (NVENC)}                                                                           & \multicolumn{1}{c}{Software (FFmpeg)}                                                                                & \multicolumn{1}{c}{Hardware (NVENC)}                                                                        & \multicolumn{1}{c}{Software (VVenC)}                                                                                 \\ \midrule
\multirow{2}{*}{P3}                                                            & \begin{tabular}[c]{@{}l@{}}VoD-HandC\\ features\end{tabular} & \begin{tabular}[c]{@{}l@{}}F5, F7, F11, F13, F26, F27, \\ F29\end{tabular}                    & \begin{tabular}[c]{@{}l@{}}F1, F5, F7, F11, F13, F17, \\ F25-F27, F29\end{tabular}                & \begin{tabular}[c]{@{}l@{}}F3, F7, F11-F13, F29, F30, \\ F27\end{tabular}                              & \begin{tabular}[c]{@{}l@{}}F1, F5, F7, F11-F13, F17, \\ F25-F27, F29\end{tabular}              & \begin{tabular}[c]{@{}l@{}}F7, F9, F11, F12, F13, F25, F27, \\ F29\end{tabular}                         \\ \cmidrule(l){2-2} \cmidrule(l){3-4} \cmidrule(l){5-6} \cmidrule(l){7-7} 
                                                                               & \begin{tabular}[c]{@{}l@{}}Live-HandC\\ features\end{tabular}     & \begin{tabular}[c]{@{}l@{}}F5, F11, F12, F13, F15, F16, \\ F34, F39\end{tabular}              & \begin{tabular}[c]{@{}l@{}}F3, F12, F13, F15, F16, F19, \\ F34\end{tabular}                       & \begin{tabular}[c]{@{}l@{}}F5, F12, F13, F15, F18, F31, \\ F33, F34, F38\end{tabular}                  & \begin{tabular}[c]{@{}l@{}}F5, F12, F13, F15, F16, F19,\\  F31, F35, F38\end{tabular}          & \begin{tabular}[c]{@{}l@{}}F1, F6, F10, F13, F15, F17, F19, \\ F33, F34, F35-F39\end{tabular}           \\ \cmidrule(l){2-2} \cmidrule(l){3-4} \cmidrule(l){5-6} \cmidrule(l){7-7}
\multirow{2}{*}{P2}                                                            & \begin{tabular}[c]{@{}l@{}}VoD-HandC\\ features\end{tabular} & \begin{tabular}[c]{@{}l@{}}F5, F7, F9-F13, F16, F25-F27,\\ F29\end{tabular}                   & \begin{tabular}[c]{@{}l@{}}F5, F7, F11, F13, F16, F25,\\  F26, F27, F29\end{tabular}              & \begin{tabular}[c]{@{}l@{}}F7, F10, F11, F13, F19, F25,\\  F27, F29\end{tabular}                       & \begin{tabular}[c]{@{}l@{}}F1, F3, F5, F7- F9, F11-F13, \\ F16, F25-F27, F29, F30\end{tabular} & F5, F7, F9-F13, F25, F27, F29                                                                           \\ \cmidrule(l){2-2} \cmidrule(l){3-4} \cmidrule(l){5-6} \cmidrule(l){7-7} 
                                                                               & \begin{tabular}[c]{@{}l@{}}Live-HandC\\ features\end{tabular}     & \begin{tabular}[c]{@{}l@{}}F6, F11-F13, F15, F16, F18, \\ F31, F33, F34, F37-F39\end{tabular} & \begin{tabular}[c]{@{}l@{}}F5, F7, F11-F13, F15, F16,\\ F18, F19, F31, F34, F38, F40\end{tabular} & \begin{tabular}[c]{@{}l@{}}F3, F4, F6, F9, F11, F13, F15, \\ F31, F34, F36, F37, F38, F40\end{tabular} & \begin{tabular}[c]{@{}l@{}}F3, F5, F11-F17, F31, F35, \\ F38, F39\end{tabular}                 & \begin{tabular}[c]{@{}l@{}}F5, F9, F10, F11, F13, F15, F21, \\ F29, F31, F32, F36, F38,F39\end{tabular} \\ \cmidrule(l){2-2} \cmidrule(l){3-4} \cmidrule(l){5-6} \cmidrule(l){7-7}
\multirow{2}{*}{P1}                                                            & \begin{tabular}[c]{@{}l@{}}VoD-HandC\\ features\end{tabular} & \begin{tabular}[c]{@{}l@{}}F3, F5-F7, F10-F13, F26, \\ F27, F29\end{tabular}                  & F5, F7, F8, F11, F13, F29                                                                         & \begin{tabular}[c]{@{}l@{}}F1-F3, F5, F7, F8, F10-F14, \\ F25, F27, F29\end{tabular}                   & \begin{tabular}[c]{@{}l@{}}F1, F3, F5-F13, F16, F25-F27, \\ F29, F30\end{tabular}              & F6-F14, F16, F19,F25, F29, F30                                                                          \\ \cmidrule(l){2-2} \cmidrule(l){3-4} \cmidrule(l){5-6} \cmidrule(l){7-7} 
                                                                               & \begin{tabular}[c]{@{}l@{}}Live-HandC\\ features\end{tabular}     & \begin{tabular}[c]{@{}l@{}}F3, F4, F9-F11, F13, F15, \\ F18, F19, F34, F38\end{tabular}       & \begin{tabular}[c]{@{}l@{}}F3-F5, F9, F10, F11, F13, \\ F15, F16, F18, F19, F38, F40\end{tabular} & \begin{tabular}[c]{@{}l@{}}F3, F5, F6, F10, F11, F13-F16,\\  F18, F38\end{tabular}                     & \begin{tabular}[c]{@{}l@{}}F1, F5, F12-F16, F35, F37, \\ F38, F39\end{tabular}                 & \begin{tabular}[c]{@{}l@{}}F5, F9, F10, F13, F18, F19, \\ F26, F32, F38, F40\end{tabular}               \\ \bottomrule
\end{tabular}%
}
\end{table*}
\subsubsection{Live-HandC features}
While VoD-HandC features have been extensively explored in computer vision and compression-related fields, their extraction is time-consuming, especially for high-resolution videos, limiting real-time applications. This is primarily due to the lack of optimization for online analysis.   In response to this issue, Menon~\etal~\cite{menon2022vca} introduced \ac{vca} as an optimized software, leveraging low-complexity features for online video content analysis, ensuring low-latency streaming and improved real-time performance. In this work, we utilize these low-complexity features, referred to as Live-HandC, for which the extraction process can be defined as follows:
\begin{equation}
\mathcal{S}_{F_i} = \Psi_{\text{Live}} (\mathcal{V}),
\end{equation}
Here, $\mathcal{S}_{F_i}$ represents the set of extracted features, $\Psi_{\text{Live}}$ corresponds to the Live-HandC feature extraction module, and $\mathcal{V}$ denotes the input video sequence. The VoD-HandC features are presented below.
\vspace{2mm}
\newline
{\bf Spatial energy (E)} is used to quantify the complexity and variation of textures within an image. It is calculated using a \ac{dct}-based energy function that determines the block-wise texture of each frame. \vspace{2mm}
\newline
{\bf Temporal energy (h)} is a temporal feature that measures changes in texture complexity and variation between consecutive frames in a video sequence. The temporal energy is calculated through the \ac{sad} of the texture energy of each frame compared to its previous frame on a block-wise basis. \vspace{2mm}
\newline
{\bf Gradient of temporal energy} ($\epsilon$) is a feature that quantifies the rate of change in texture complexity and variation between consecutive frames in a video sequence. It is calculated as the ratio of the difference between \ac{h} values of the $(p-1)^{th}$ and $p^{th}$ frames with the \ac{h} value of the $(p-1)^{th}$ frame. \vspace{2mm}
\newline
{\bf Brightness (BR)} is a spatial feature that quantifies the overall light intensity of a frame. \vspace{2mm}

Finally, Table~\ref{VCA_feat} provides the ~full set of listed features and their statistics. To provide a comprehensive understanding of the entire video,  various statistics, such as mean, standard deviation, maximum, minimum, $25^{th}$ percentile, $50^{th}$ percentile, $75^{th}$ percentile, \ac{iqr}, skewness and kurtosis are computed for each feature.

\subsubsection{Feature selection}

\begin{figure}[!t]
\centering
\scriptsize
\centering
\begin{minipage}[b]{0.48\linewidth}
\centering
\centerline{\includegraphics[width=0.95\linewidth]{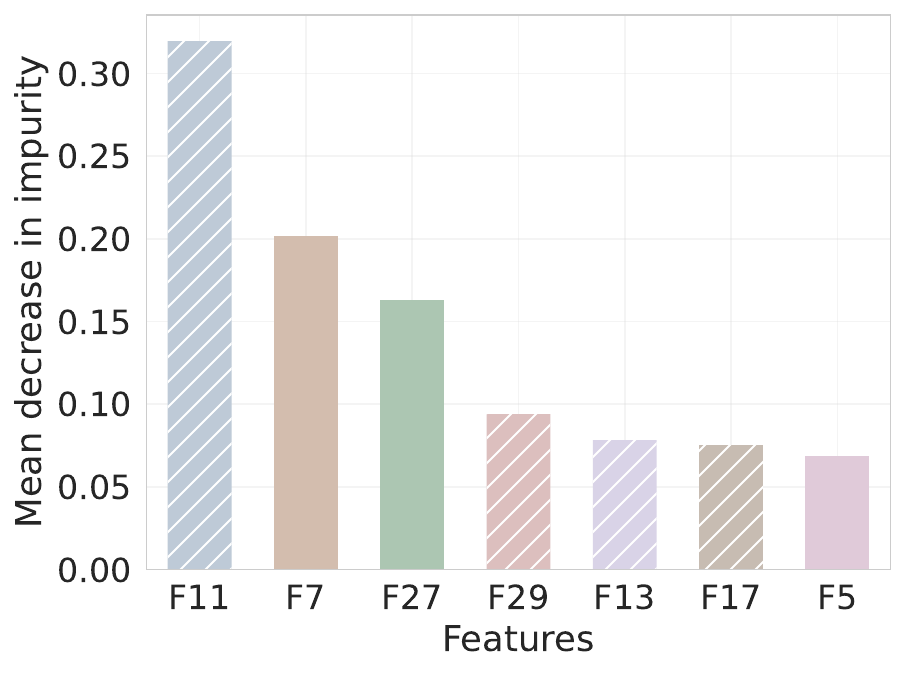}}
(a) VoD-HandC feature importance
\end{minipage}
\begin{minipage}[b]{0.48\linewidth}
\centering
\centerline{\includegraphics[width=0.95\linewidth]{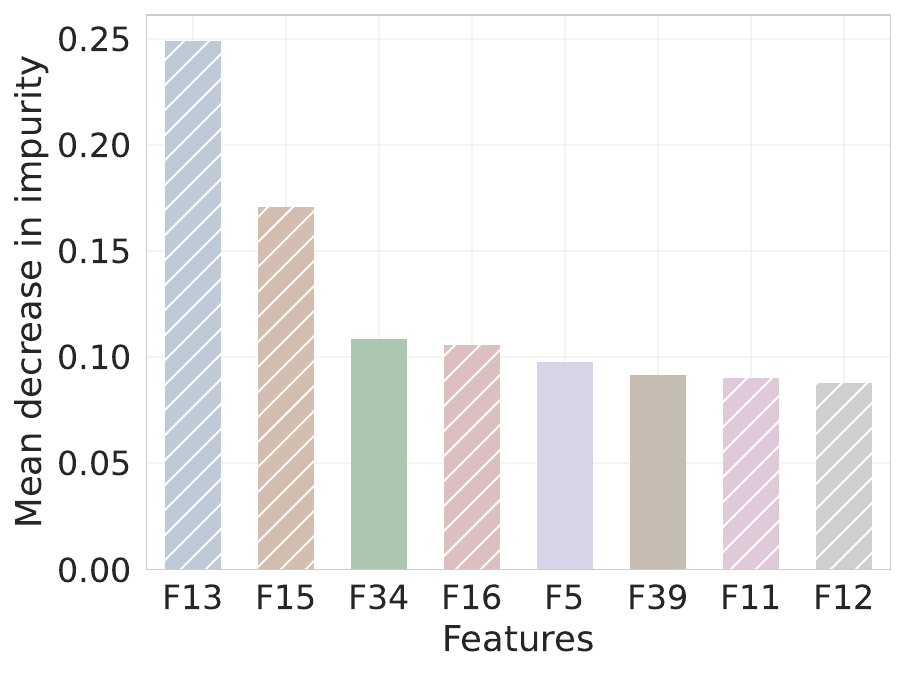}}
(b) Live-HandC feature importance
\end{minipage}

\caption{Feature importance of selected features for P3 cross-point prediction for x264 software encoding in terms of YPSNR. Dashed bars represent temporal features.}
\label{features_impor}
\end{figure}

\begin{figure}[!]
\centering
\scriptsize
\centering
\begin{minipage}[b]{0.48\linewidth}
\centering
\centerline{\includegraphics[width=0.92\linewidth]{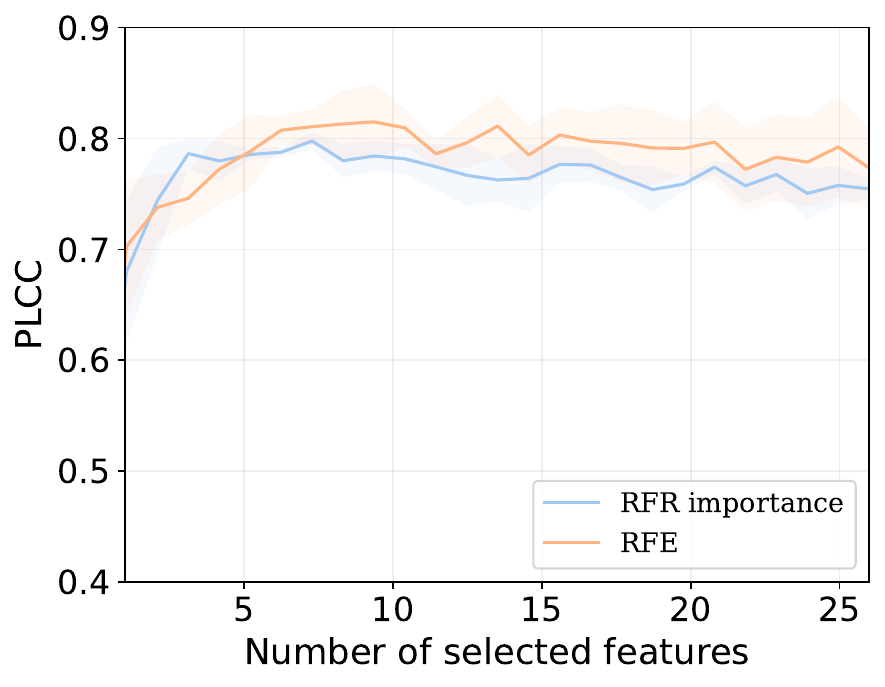}}
(a) VoD-HandC feature selection
\end{minipage}
\begin{minipage}[b]{0.48\linewidth}
\centering
\centerline{\includegraphics[width=0.96\linewidth]{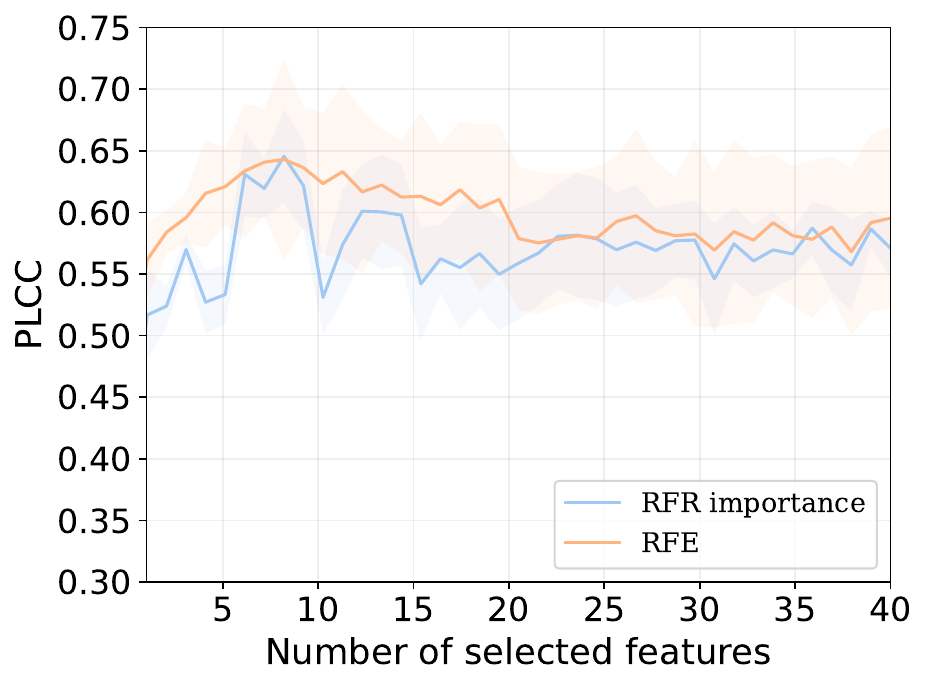}}
(b) Live-HandC feature selection
\end{minipage}
\caption{Feature selection performance (PLCC) for P3 cross-point prediction for x264 software encoding in terms of YPSNR. The shaded error bar represents the standard deviation of PLCC over three iterations.}
\vspace{-5mm}
\label{features_selec}
\end{figure}

In the context of increasing data dimensionality, determining which features to incorporate into machine learning models is crucial. Feature selection addresses this challenge by identifying the most pertinent and informative features within a dataset, thereby (\textit{i}) enhancing models by reducing dimensionality, (\textit{ii}) preventing overfitting, and (\textit{iii}) lowering computational complexity. In our approach, we employ two types of feature selection algorithms to distill the initial feature set, denoted as $\mathcal{S}{F_i}$, into a refined feature set, referred to as $\mathcal{S}{F_s}$, for both VoD-HandC and Live-HandC features. This process ensures optimal model efficiency and accuracy. The first type of feature selection employs model-based feature selectors. In this category, we utilize the \ac{rfr} algorithm, known for its robustness and effectiveness in handling complex datasets. The \ac{rfr} algorithm fits a regression model, capturing the relationships between features and the target variable. One key advantage of \ac{rfr}-based feature selectors is their capacity to calculate the permutation importance of each feature, as illustrated in Fig.~\ref{features_impor}. The feature importance is measured using the mean decrease in impurity, which quantifies how much each feature contributes to reducing uncertainty in the model. Higher impurity reduction indicates a more influential feature. By ranking the features based on their permutation importance, we can identify and eliminate the least significant ones. The second technique involves applying \ac{rfe} to remove the weakest features systematically. The \ac{rfe} operates by iteratively fitting a machine learning model, ranking features based on their importance, and subsequently eliminating the least important ones. In this study, we utilize ExtraTrees~\cite{geurts2006extremely} as the target regressor. In the first step, we evaluated both of these feature selection methods on three test-training iterations using stratified sampling.  Fig.~\ref{features_selec} illustrates the median \ac{plcc} performance for P3 cross point prediction for x264 software encoding in terms of YPSNR. Following this comparison, we proceeded with the \ac{rfe} technique due to its superior results. The resulting selected features for predicting cross-over bitrates are presented in Table~\ref{YPSNR_selec} for \ac{ypsnr} \ac{rd} curves. Finally, these selected features are used to train a regression machine learning model for predicting the convex hull, as outlined below:
\begin{equation}
\hat{P_{k}} = \theta_{k} (\mathcal{S}_{F_s}), 
\end{equation}
where $\hat{P_{k}}$ denotes the predicted value for $P_{k}$ on the bitrate ladder $P=(P_{1}, P_{2}, P_{3})$, $\theta_{k}$ is the parametric function of the regression \ac{ml} model, while $\mathcal{S}_{F_s} = \{\text{F1}, \dots  \text{Fm}\} $ represents the selected feature set.

\subsection{Methods based on deep neural network}
\label{dl:models}

\begin{figure*}[!]
  \centering
  \centerline{\includegraphics[width=0.83\linewidth]{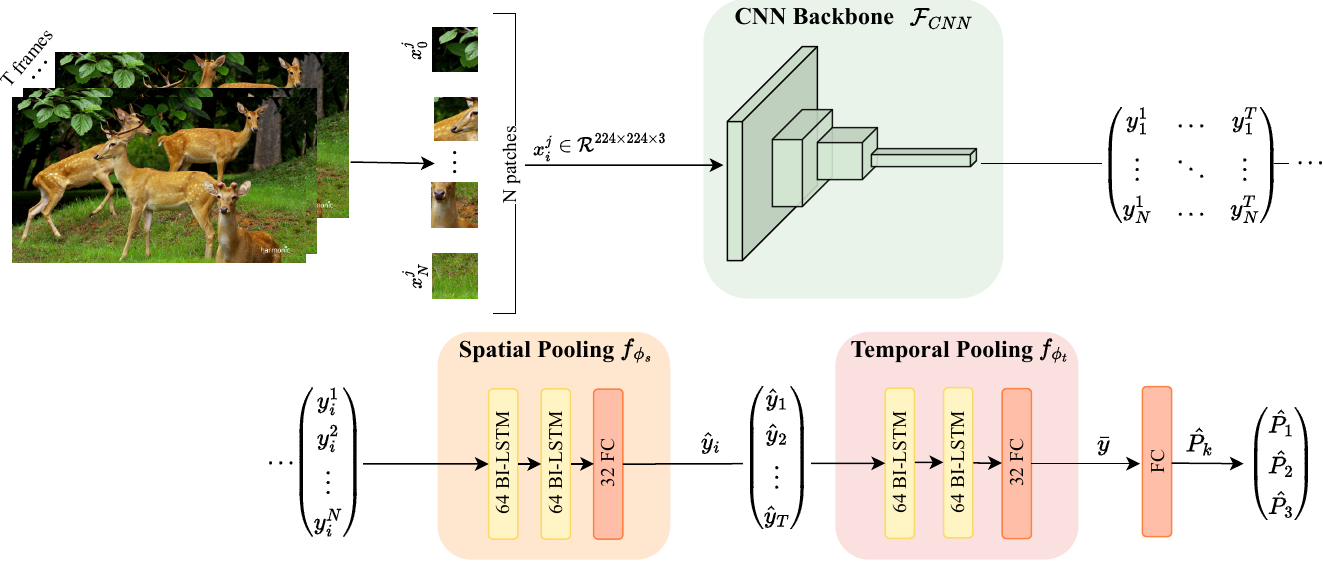}}
\caption{The overall framework of the proposed DNN methods. The feature extraction module extracts spatial features $\bm{y}_i^j$ from patches $\bm{x}_i^j$. The spatial and temporal pooling modules aggregate features into a final vector $\bar{\bm{y}}$. Finally, the regression module uses the final vector $\bar{\bm{y}}$ to predict the cross-over bitrates $\hat{P_{k}}$.}
\vspace{-5mm}
\label{overall_dnn}
\end{figure*}

Recently, \acp{dnn} have gained tremendous attention in computer vision due to their remarkable abilities to process complex visual data and solve a wide range of computer vision problems. In light of these advances, exploring the merit of using deep learning techniques to construct a content-gnostic bitrate ladder in the adaptive video streaming field is well-deserved. Thus, we investigate the application of \ac{dnn} to achieve this goal in this subsection. As illustrated in Fig.~\ref{overall_dnn}, the proposed  \ac{dnn} solution's framework consists of four main modules: feature extraction, spatial pooling, temporal pooling, and convex hull regression module.

\subsubsection{Feature extraction}

Deep \ac{cnn} features offer a more powerful approach than traditional \ac{ml}-based methods by automatically learning hierarchical representations of local features, enabling the extraction of higher-level abstractions from raw video input. However, the efficacy of \acp{cnn} is closely tied to the amount of training data available, and the proposed dataset is still much smaller than the typical computer vision datasets with millions of samples. To enhance feature representation, we used pre-trained models on the extensive ImageNet dataset \cite{deng2009imagenet} as backbone models, serving as deep feature descriptors. This leverages the knowledge learned from vast amounts of training data, significantly improving overall the performance of \ac{dnn} approaches in the context of convex hull prediction. 

When using pre-trained models, one challenge is dealing with their fixed input shape requirements. Two potential alternative solutions to address this constraint are resizing the input frame or dividing it into multiple patches. However, the first method could adversely lead to a loss of fine-grained details and introduce distortions, which may affect the network's ability to extract meaningful features for the accurate prediction of bitrate-resolution pairs.  Consequently, we chose the second approach.

Let us consider the input sequence $\mathcal{V}$ as a set of $T$ consecutive frames: $\mathcal{V} = \{\bm{x}_1, \bm{x}_2,  \dots, \bm{x}_T \}$. For each frame $\bm{x}_{i}$ a sliding window is used to extract $N$ non-overlapping patches $\bm{x}_i^j$ of size $224\times224$, where $j \in \{1, \dots, N \}$ and $i \in \{1, \dots, T \}$. Then, these patches $\bm{x}_i^j$ are fed into the \ac{cnn} backbone pre-trained on ImageNet \cite{deng2009imagenet} for the extraction of spatial features $\bm{y}_i^j$ expressed as follows:
\begin{equation}
\bm{y}_i^j = \mathcal{F}_{CNN} (\bm{x}_i^j).
\end{equation}
where $\mathcal{F}_{CNN}$ denotes the \ac{cnn} backbone. 

\subsubsection{Spatial pooling}

after extracting features from each patch, they are aggregated into a single vector per frame, encapsulating the frame's spatial information. Traditionally done using spatial pooling, we instead employ a \ac{bi-lstm} network due to the sequential and correlated nature of patches. This approach allows the model to process features in both directions, effectively summarizing local features, reducing computational complexity, and efficiently capturing short-term dependencies between neighboring patches. 

The module is composed of two \ac{bi-lstm} layers followed by a \ac{fc} layer with 256 nodes. We have found that using this architecture leads to the best results in our experiments. The feature vector $(\bm{y}_i^1, \bm{y}_i^2,  \dots, \bm{y}_i^N)^T$ of a frame $i$ is fed into the spatial pooling module, expressed as:
\begin{equation}
    \hat{\bm{y}}_i = f_{\phi_s} (\bm{y}_i^1 , \bm{y}_i^2, \dots  , \bm{y}_i^N),\hspace*{3mm} \forall i \in \{1, \dots, T \},
\end{equation}
where $f_{\phi_s}$ is the parametric function of the spatial pooling module with the training parameters $\phi_s$.

\subsubsection{Temporal pooling}
parallel to the spatial aggregation process, temporal aggregation of frame features into a single vector per sequence is also a critical step. This temporal modeling module captures the dynamic changes and patterns over time, providing a crucial dimension of information for predicting the bitrate ladder.
Hence, we employ a temporal modeling module that utilizes a \ac{bi-lstm} network to consolidate the frame-level features, denoted as $(\hat{\bm{y}}_1 , \hat{\bm{y}}_2, \dots , \hat{\bm{y}}_T)$, into a comprehensive global feature vector $\bar{\bm{y}}$ that encapsulates the entire video sequence. \Ac{bi-lstm} networks are adept at considering both forward and backward information, enabling the capture of long-range dependencies among frames. Analogous to the spatial modeling, this module comprises two \ac{bi-lstm} layers, each with 64 cells, and is subsequently followed by a \ac{fc} layer containing 256 nodes. The temporal modeling module can be expressed as follows:
\begin{equation}
    \bar{\bm{y}} = f_{\phi_t} (\hat{\bm{y}}_1 , \hat{\bm{y}}_2, \dots , \hat{\bm{y}}_T),
\end{equation}
where $f_{\phi_t}$ is the parametric function of the temporal modeling module with the training parameters $\phi_t$.

\subsubsection{Convex hull regression}
following the extraction and aggregation of deep features into a single representative vector $\bar{y}$, the challenge lies in mapping these features onto the respective point in the final convex hull. To address this, we employed one node in a \ac{fc} layer as a regression model with a linear activation function to predict a single point on the convex hull. To predict the complete bitrate ladder, $P=(P_{1}, P_{2}, P_{3})$,  we implemented three distinct models, each consisting of a spatial and temporal module as described above, followed by the regression module.  We have found that using this architecture, with separate models for each point on the convex hull, leads to the best results in our experiments.  Consequently, the prediction for each point in the convex hull can be formulated as follows: 

\begin{equation}
\hat{P_{k}} = \zeta_{k} (\bar{\bm{y}}),
\end{equation}
where $\hat{P_{k}}$ represents the predicted value for $P_{k}$ on the bitrate ladder $P=(P_{1}, P_{2}, P_{3})$ and $\zeta_{k}$ denotes the \ac{fc} layer, where $k \in \{1, 2, 3 \}$.

The training process is performed with 200 epochs using Adam optimizer~\cite{DBLP:journals/corr/KingmaB14} with an initial learning rate of $1e-4$, bat ch size of 16 and the \ac{mse} as loss function.  All backbones are frozen during this process, and the rest of the network is trained.

\section{EXPERIMENTAL RESULTS}
\label{results}

In this section, we first define the experimental setup, including the baselines, the evaluation methods, and the implementation details. Then, we present the coding performance in terms of coding efficiency and complexity.

\subsection{Experimental setup}
\begin{table*}[]
\centering
\caption{Performance comparison in YPSNR / VMAF of the considered models on the proposed dataset with AVC encoder. The top result for each encoder type is highlighted in boldface.}
\label{avc_result}
\resizebox{\textwidth}{!}{%
\begin{tabular}{@{}lllllccc@{}}
\toprule
\multicolumn{2}{l}{Encoder type}        & \multicolumn{6}{c}{AVC software encoding}                                    \\ \cmidrule(l){1-8} 
\multicolumn{2}{l}{Model \textbackslash{} Metric}       & \multicolumn{1}{c}{R2 $\uparrow$} & \multicolumn{1}{c}{SROCC $\uparrow$} & \multicolumn{1}{c}{PLCC $\uparrow$} & Accuracy $\uparrow$       & BD-BR vs EEL $\downarrow$       & BD-BR vs SL $\downarrow$          \\ \midrule
\multirow{4}{*}{\rotatebox[origin=c]{90}{\scriptsize VoD-HandC}} & ExtraTrees        & \textbf{0.7265} / \textbf{0.5559}        & \textbf{0.7466} / \textbf{0.7252}           & \textbf{0.8689} / 0.7741          & \textbf{0.8462} / 0.8056 & \textbf{2.529\%} / \textbf{4.193\%}  & \textbf{-6.785\%} / \textbf{-7.965\%}  \\
& Random Forests    & 0.6760 / 0.5360        & 0.7114 / 0.6984           & 0.8308 / \textbf{0.7751}          & 0.8309 / 0.7932 & 2.655\% / 4.229\%  & -6.852\% / -7.454\%  \\
& XGBoost  & 0.6629 / 0.5385        & 0.7263 / 0.6418           & 0.8241 / 0.6241          & 0.8395 / 0.7984 & 2.749\% / 4.326\%  & -6.893\% /  -7.863\%   \\
&   LightGBM       & 0.6143 / 0.5192        & 0.6472 / 0.6318           & 0.7925 / 0.7341          & 0.8235 / \textbf{0.8086} & 2.850\% / 4.465\%  & -5.719\% / -6.457\%  \\    \addlinespace[0.3ex] \hdashline \addlinespace[0.3ex]
\multirow{4}{*}{\rotatebox[origin=c]{90}{\scriptsize Live-HandC}} & ExtraTrees          & 0.3854 / 0.2869        & 0.6885 / 0.6209           & 0.6347 / 0.5671          & 0.7980 / 0.7558 & 4.040\% / 5.226\%  & -5.307\% / -6.029\%  \\
& Random Forests      & 0.3325 / 0.2429        & 0.6629 / 0.6072           & 0.5788 / 0.4399          & 0.7977 / 0.7589 & 4.200\% / 6.057\%  & -5.229\% / -5.253\% \\
& XGBoost      & 0.3395 / 0.2417        & 0.6700 / 0.5905           & 0.5864 / 0.4951          & 0.8009 / 0.7483 & 3.718\% / 5.935\%  & -5.401\% / -7.123\%  \\     
& LightGBM    & 0.2729 / 0.1745        & 0.5993 / 0.5114           & 0.5338 / 0.4327          & 0.7831 / 0.7350 & 5.313\% / 7.394\%  & -3.865\% / -3.788\%  \\ \addlinespace[0.3ex] \hdashline \addlinespace[0.3ex]
\multirow{4}{*}{\rotatebox[origin=c]{90}{\scriptsize Deep features}} & DensNet169         & 0.4964 / 0.5294        & 0.5487 / 0.6013           & 0.6813 / 0.7507          & 0.8032 / 0.7913 & 3.582\% / 5.264\%  & -5.174\% / -6.574\%   \\
& VGG16              & 0.5119 / 0.5133        & 0.5295 / 0.6000           & 0.7125 / 0.7373          & 0.7908 / 0.7845 & 3.368\% / 7.966\%  & -5.165\% / -6.637\%  \\
& ResNet-50         & 0.5416 / 0.5421        & 0.5371 / 0.6438           & 0.7425 / 0.7437          & 0.8155 / 0.7988 & 2.901\% / 4.268\%  & -6.694\% / -7.025\%  \\
& ConvNeXtBase       & 0.5381 / 0.5384        & 0.5544 / 0.5789           & 0.7323 / 0.7416          & 0.7982 / 0.7841 & 3.323\% / 5.678\%  & -5.489\% / -6.051\%  \\ \addlinespace[1.05ex]\midrule
\multicolumn{2}{l}{Encoder type}         & \multicolumn{6}{c}{AVC hardware encoding}                                                                                                   \\ \cmidrule(l){1-8} 
\multicolumn{2}{l}{Model \textbackslash{} Metric}       & \multicolumn{1}{c}{R2 $\uparrow$} & \multicolumn{1}{c}{SROCC $\uparrow$} & \multicolumn{1}{c}{PLCC $\uparrow$} & Accuracy  $\uparrow$      & BD-BR vs EEL $\downarrow$       & BD-BR vs SL $\downarrow$         \\ \midrule
\multirow{4}{*}{\rotatebox[origin=c]{90}{\scriptsize VoD-HandC}} & ExtraTrees        & \textbf{0.6461} / 0.4431        & 0.5994 / 0.5451           & \textbf{0.8200} / \textbf{0.6885}          & \textbf{0.8075} / \textbf{0.7856} & \textbf{3.017\%} / \textbf{3.478\%}  & \textbf{-6.439\%} / \textbf{-5.594\%}  \\
& Random Forests      & 0.5749 / 0.3289        & 0.5568 / 0.5152           & 0.7824 / 0.6343          & 0.7958 / 0.7719 & 3.502\% / 4.251\%  & -5.979\% / -4.938\%   \\
&  LightGBM        & 0.5591 / \textbf{0.4706}        & 0.5097 / 0.4561           & 0.7752 / 0.6231          & 0.7871 / 0.7565 & 4.343\% / 4.663\%  & -5.642\% / -4.856\%  \\
& XGBoost    & 0.5117 / 0.3207        & 0.5391 / 0.4696           & 0.7385 / 0.6073          & 0.7849 / 0.7507 & 4.452\% / 5.675\%  & -3.982\% / -4.041\%  \\ \addlinespace[0.3ex] \hdashline \addlinespace[0.3ex]
\multirow{4}{*}{\rotatebox[origin=c]{90}{\scriptsize Live-HandC}} & ExtraTrees     & 0.2943 / 0.2617        & 0.5092 / 0.4290           & 0.5787 / 0.5438          & 0.7622 / 0.7432 & 4.547\% / 4.606\%  & -5.292\% / -4.512\%  \\ 
& Random Forests    & 0.2733 / 0.3032        & 0.4975 / 0.4311           & 0.5451 / 0.5709          & 0.7540 / 0.7347 & 5.523\% / 4.892\%  & -4.352\% / -3.952\%  \\  
&       LightGBM    & 0.2560 / 0.2418        & 0.4647 / 0.3657           & 0.5124 / 0.5225          & 0.7589 / 0.7492 & 5.363\% / 5.453\%  & -4.605\% / -4.152\%  \\
& XGBoost        & 0.1997 / 0.1926        & 0.4075 / 0.3524           & 0.4784 / 0.4650          & 0.7543 / 0.7312 & 5.920\% / 6.253\%  & -3.105\% / -3.573\%  \\ \addlinespace[0.3ex] \hdashline \addlinespace[0.3ex]
\multirow{4}{*}{\rotatebox[origin=c]{90}{\scriptsize Deep features}} & DensNet169         & 0.3875 / 0.3880        & 0.5986 / 0.5275           & 0.6550 / 0.6592          & 0.7890 / 0.7811 & 3.497\% / 3.997\%  & -5.512\% / -4.248\%  \\
& VGG16               & 0.4477 / 0.3849        & 0.6023 / 0.5095           & 0.6871 / 0.6431          & 0.7921 / 0.7663 & 3.678\% / 3.999\%  & -5.041\% / -3.812\%  \\
& ResNet-50           & 0.5534 / 0.4152        & \textbf{0.6048} / \textbf{0.5567}           & 0.7042 / 0.6462          & 0.7940 / 0.7803 & 3.249\% / 3.807\%  & -5.754\% / -4.502\%  \\
& ConvNeXtBase        & 0.5116 / 0.4033        & 0.5601 / 0.4737           & 0.7022 / 0.6349          & 0.7967 / 0.7716 & 3.320\% / 3.932\%  & -5.605\% / -4.277\%  \\ \addlinespace[.55ex] \bottomrule
\end{tabular}%

}\vspace{-5mm}

\end{table*}
\subsubsection{Baselines}
due to the lack of publicly available implementations for many of the methods discussed in Section~\ref{sec:review}, we opted to benchmark the performance of our proposed models against two alternative methods. These methods, outlined below, offer a trade-off between computational cost, simplicity, and compression performance:  \vspace{-1mm}\\\\
{\bf \Ac{eel}}: this approach, based on exhaustive encoding detailed in Subsection~\ref{convex_construction},  provides a reference point for our performance measurements by generating a fully specialized convex hull for each content. The \ac{eel} approach also serves as the ground truth during the training stage. \\
{\bf\Ac{sl}}: this method creates a fixed bitrate resolution pair by averaging the ground truth bitrate ladders obtained from the training dataset (using the \ac{eel} approach). 

\subsubsection{Considered models} In our benchmark, we consider two sets of handcrafted features and four deep features extracted from popular deep backbone \ac{cnn} models (i.e., excluding the classification head): VGG16 (14.71M parameters)~\cite{zhang2015accelerating}, DenseNet169 (12.64M parameters)~\cite{huang2017densely}, ResNet-50 (23.59M parameters)~\cite{he2016deep}, and ConvNeXtBase (87.57M parameters)~\cite{liu2022convnet}.s
The two sets of handcrafted features, VoD-HandC and Live-HandC,  are trained with four machine-learning regression algorithms, including ExtraTrees Regressor~\cite{geurts2006extremely}, Random Forest~\cite{ho1995random}, XGBoost~\cite{chen2015xgboost}, and LightGBM~\cite{ke2017lightgbm}, resulting in eight handcrafted-based models. The deep features are processed with the spatial and temporal pooling models described in Section~\ref{dl:models}, resulting in four \ac{dl}-based methods.

\subsubsection{Evaluation metrics}
for a comprehensive evaluation, we split the \ac{avsd} dataset into two non-overlapping subsets, 80\% for training and the remaining 20\% for testing. To minimize sampling bias and ensure the representation of all types of content, we employed stratified sampling, repeating this process three times to mitigate the impact of sampling errors. 
To assess the performance of the considered models, we used three correlation metrics, including \ac{r2}, \ac{srocc}, and \ac{plcc}, to measure the correlation between the predicted cross-over bitrates and the reference cross-over bitrates generated by the EEL approach. We also use the accuracy to evaluate the performance of each approach in predicting the optimal resolution over all tested bitrates. Finally, we compute the \ac{bd-br} score between the predicted and the reference bitrate ladders, giving the rate gain or loss in percentage of each method compared to the anchor. This is done by encoding video sequences at various bitrates, with their resolutions defined by both ladders, and then using the resulting bitrate and quality values to determine the \ac{bd-br}.

\begin{table*}[]
\centering
\caption{Performance comparison in YPSNR / VMAF of the considered models on the proposed dataset with HEVC encoder. The top result for each encoder type is highlighted in boldface.}
\label{hevc_result}
\resizebox{\textwidth}{!}{%
\begin{tabular}{@{}lllllccc@{}}
\toprule
\multicolumn{2}{l}{Encoder type}        & \multicolumn{6}{c}{HEVC software encoding}                                    \\ \cmidrule(l){1-8} 
\multicolumn{2}{l}{Model \textbackslash{} Metric}       & \multicolumn{1}{c}{R2 $\uparrow$} & \multicolumn{1}{c}{SROCC $\uparrow$} & \multicolumn{1}{c}{PLCC $\uparrow$} & Accuracy $\uparrow$       & BD-BR vs EEL $\downarrow$       & BD-BR vs SL $\downarrow$          \\ \midrule
\multirow{4}{*}{\rotatebox[origin=c]{90}{\scriptsize VoD-HandC}} & ExtraTrees        & \textbf{0.6479} / \textbf{0.4929}        & \textbf{0.6730} / \textbf{0.6286}           & \textbf{0.8154} / \textbf{0.7131}          & \textbf{0.8852} / \textbf{0.8685} & \textbf{1.776\%} / \textbf{2.545\%}  & \textbf{-5.997\%} / \textbf{-5.583\%}  \\
& Random Forests    & 0.6079 / 0.4293        & 0.6452 / 0.6087           & 0.7847 / 0.6679          & 0.8814 / 0.8626 & 2.365\% / 2.670\% & -5.886\% / -5.721\%  \\
&  XGBoost   & 0.5659 / 0.3877        & 0.6312 / 0.5867           & 0.7767 / 0.6541          & 0.8775 / 0.8606 & 2.909\% / 3.331\%  & -5.524\% / -4.943\%  \\
& LightGBM         & 0.5487 / 0.4095        & 0.6436 / 0.5762           & 0.7518 / 0.6496          & 0.8665 / 0.8397 & 2.023\% / 3.841\%  & -5.293\% / -4.785\%  \\    \addlinespace[0.3ex] \hdashline \addlinespace[0.3ex]
\multirow{4}{*}{\rotatebox[origin=c]{90}{\scriptsize Live-HandC}} & ExtraTrees          & 0.2700 / 0.3250        & 0.4941 / 0.5341           & 0.5411 / 0.5832          & 0.8354 / 0.8396 & 3.448\% / 4.037\%  & -4.805\% / -5.179\%  \\
& Random Forests   &    0.2627 / 0.2301        & 0.5083 / 0.5363           & 0.5303 / 0.5060          & 0.8329 / 0.8303 & 3.870\% / 4.310\%  & -4.784\% / -4.803\% \\
& XGBoost      & 0.2507 / 0.1920        & 0.5181 / 0.4978           & 0.5317 / 0.4471          & 0.8303 / 0.8299 & 3.866\% / 4.041\%  & -4.828\% / -4.585\%  \\     
& LightGBM    & 0.2438 / 0.1687        & 0.5172 / 0.4279           & 0.5088 / 0.4169          & 0.8315 / 0.8149 & 3.840\% / 4.188\%  & -4.889\% / -4.531\%  \\ \addlinespace[0.3ex] \hdashline \addlinespace[0.3ex]
\multirow{4}{*}{\rotatebox[origin=c]{90}{\scriptsize Deep features}} & DensNet169         & 0.3306 / 0.3139        & 0.3778 / 0.5374           & 0.5703 / 0.5250          & 0.8793 / 0.8395 & 3.300\% / 3.181\%  & -4.915\% / -4.823\%   \\
& VGG16               & 0.2787 / 0.2511        & 0.3832 / 0.5531           & 0.5767 / 0.5083          & 0.8637 / 0.8227 & 3.378\% /  3.329\% & -4.845\% / -4.751\%  \\
& ResNet-50         & 0.5652 / 0.3204        & 0.5057 / 0.5802           & 0.6349 / 0.5400          & 0.8805 / 0.8539 & 2.755\% / 2.999\%  & -5.387\% / -5.012\%  \\
& ConvNeXtBase       & 0.3188 / 0.2321        & 0.4788 / 0.4650           & 0.5884 / 0.5005          & 0.8756 / 0.8252 & 3.600\% / 3.434\%  & -4.892\% / -4.712\%  \\ \addlinespace[.55ex] \midrule
\multicolumn{2}{l}{Encoder type}         & \multicolumn{6}{c}{HEVC hardware encoding}                                                                                                   \\ \cmidrule(l){1-8} 
\multicolumn{2}{l}{Model \textbackslash{} Metric}       & \multicolumn{1}{c}{R2 $\uparrow$} & \multicolumn{1}{c}{SROCC $\uparrow$} & \multicolumn{1}{c}{PLCC $\uparrow$} & Accuracy  $\uparrow$      & BD-BR vs EEL $\downarrow$       & BD-BR vs SL $\downarrow$         \\  \midrule
\multirow{4}{*}{\rotatebox[origin=c]{90}{\scriptsize VoD-HandC}} & ExtraTrees        & \textbf{0.4728} / 0.3822        & 0.5927 / 0.4424           & \textbf{0.7027} / 0.6452          & \textbf{0.8373} / 0.7815 & \textbf{2.644\%} / 4.850\%  & -5.319\% / -5.068\%  \\
& Random Forests      & 0.4113 / 0.3481        & 0.5475 / 0.3934           & 0.6527 / 0.6057          & 0.8278 / 0.7800 & 3.292\% / 4.724\%  & -4.896\% / -4.932\%    \\
& XGBoost       & 0.4139 / 0.2794        & 0.5219 / 0.3452           & 0.6536 / 0.5462          & 0.8228 / 0.7669 & 3.880\% / 5.370\%  & -4.916\% / -4.472\%  \\
& LightGBM    & 0.3039 / 0.1876        & 0.5304 / 0.3347           & 0.6428 / 0.4816          & 0.8029 / 0.7371 & 4.501\% / 5.915\%  & -4.922\% / -3.728\%  \\ \addlinespace[0.3ex] \hdashline \addlinespace[0.3ex]
\multirow{4}{*}{\rotatebox[origin=c]{90}{\scriptsize Live-HandC}} & ExtraTrees     & 0.3188 / 0.2321        & 0.4788 / 0.4650           & 0.5884 / 0.5005          & 0.7894 / 0.7715 & 4.610\% / 5.100\%  & -4.209\% / -3.533\%   \\ 
& Random Forests    & 0.2854 / 0.1866        & 0.4879 / 0.3989           & 0.5612 / 0.4020          & 0.7841 / 0.7613 & 5.205\% / 5.921\%  & -3.951\% / -3.258\%  \\  
&   XGBoost        & 0.2490 / 0.1391        & 0.4843 / 0.3629           & 0.5293 / 0.4686          & 0.7793 / 0.7627 & 5.049\% / 5.930\%  & -4.094\% / -2.807\%  \\
& LightGBM        & 0.3039 / 0.1876        & 0.5304 / 0.3347           & 0.6428 / 0.4816          & 0.8029 / 0.7371 & 4.501\% / 5.915\%  & -4.922\% / -3.728\%  \\ \addlinespace[0.3ex] \hdashline \addlinespace[0.3ex]
\multirow{4}{*}{\rotatebox[origin=c]{90}{\scriptsize Deep features}} & DensNet169         & 0.3750 / \textbf{0.4408}        & 0.5710 / \textbf{0.5477}           & 0.6109 / \textbf{0.7133}          & 0.8272 / \textbf{0.8143} & 3.059\% / 4.333\%  & -4.967\% / -5.018\%  \\
& VGG16               & 0.3422 / 0.3559        & 0.5584 / 0.4899           & 0.6159 / 0.6239          & 0.7649 / 0.7529 & 3.750\% / 5.335\%  & -4.807\% / -3.931\%  \\
& ResNet-50           & 0.3466 / 0.4229        & \textbf{0.5965} / 0.5251           & 0.6354 / 0.7025          & 0.8278 / 0.8003 & 2.811\% / \textbf{4.150\%}  & \textbf{-5.139\%} / \textbf{-5.533\%}  \\
& ConvNeXtBase        & 0.2440 / 0.3490        & 0.5013 / 0.5127           & 0.5284 / 0.6253          & 0.7420 / 0.7544 & 3.916\% / 5.417\%  & -4.745\% /  -3.590\%  \\ \addlinespace[.55ex] \bottomrule
\end{tabular}%

} 
\vspace{-5mm}
\end{table*}

\subsection{Coding performance}

Tables~\ref{avc_result},~\ref{hevc_result}, and~\ref{vvc__result} give the median performance of the considered models over three stratified iterations on the \ac{avsd} dataset, using the \ac{avc}, \ac{hevc}, and \ac{vvc} encoders, respectively for both \ac{ypsnr} and \ac{vmaf} objective quality metrics.

\begin{figure}[t]
\centering
 \includegraphics[width=0.94\linewidth]{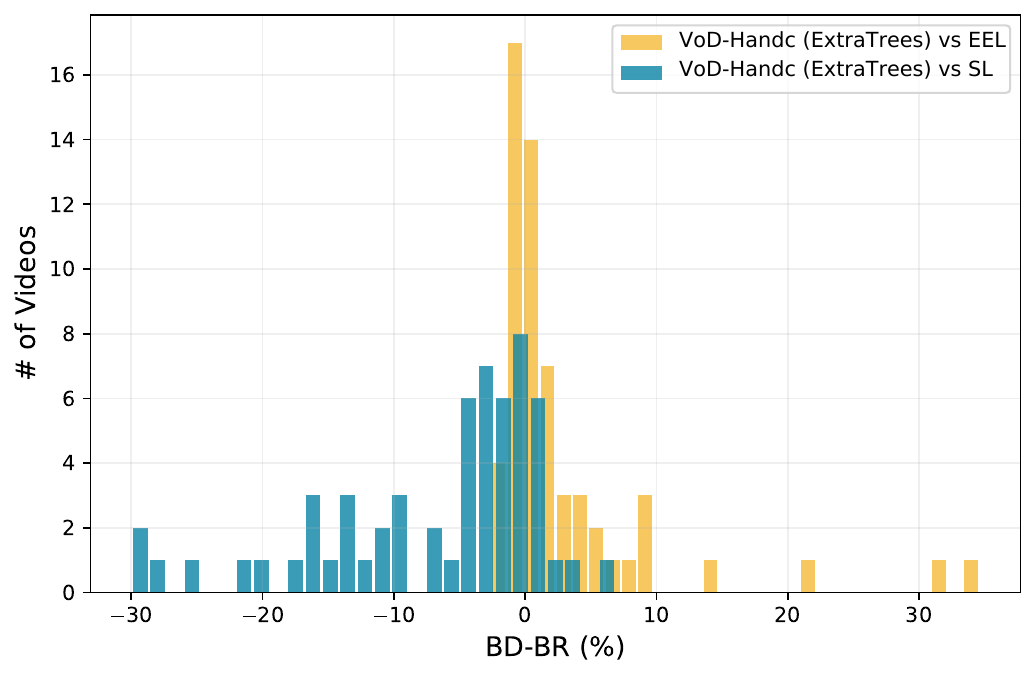}
\caption{Histogram of BD-BR for VoD-HandC ExtraTrees model compared to EEl and RL approaches for AVC softwaer encoding in terms of YPSNR.}
\label{bd_br_fig} 
\end{figure}

Table~\ref{avc_result} shows the performance of these models using \ac{avc} software and hardware encoders. The VoD-HandC features-based models consistently achieve high scores across multiple metrics in both encoding scenarios. For instance, the ExtraTrees Regressor model achieves an average accuracy of 88\%/86\% in predicting cross-over bitrates, resulting in a gain of 5.99\%/5.58\% over the static approach at the cost of a slight \ac{bd-br} loss of 1.77\%/2.54\% compared to the \ac{eel} method using software encoding, in terms of \ac{ypsnr}/\ac{vmaf}. The Live-HandC features-based models (\ac{vca} features), although not as strong as the VoD-HandC features, still outperform the \ac{sl} approach, resulting in an average \ac{bd-br} gain of 4.95\%/5.54\%  with software encoding and 4.33\%/4.04\% with hardware encoding, in terms of \ac{ypsnr}/\ac{vmaf}. On the other hand, \ac{dnn}-based models show relatively competitive performance compared to the VoD-HandC features-based models, particularly ResNet-50, which achieved the best results among \ac{cnn} baselines in both software and hardware encoding scenarios. Fig.~\ref{bd_br_fig} illustrates the histogram of \ac{bd-br} per tested sequence compared to \ac{eel} and \ac{sl} methods. The figure shows that the bitrate gain is content dependent, ranging from -30\% gain to a slight loss for a few sequences (outliers) compared to the \ac{sl} method.

Table~\ref{hevc_result} presents the performance comparison using the \ac{hevc} encoders. The VoD-HandC feature-based (VoD-HandC) models exhibit strong performance in \ac{hevc} software encoding, notably with the ExtraTrees Regressor model achieving the highest scores in terms of \ac{ypsnr} and \ac{vmaf}. It outperforms the static approach with an overall gain of 5.99\%/5.58\%  while incurring only a slight loss of 1.77\%/2.54\% compared to the \ac{eel} method in terms of \ac{bd-br}. Additionally, the Live-HandC feature-based (Live-HandC) models perform consistently better than the static approach \ac{sl},  particularly the ExtraTrees Regressor, with an average \ac{bd-br} gain of 4.80\%/5.17\% using software encoding and 4.20\%/3.53\%  using hardware encoding. In comparison, the \ac{dnn}-based models, particularly ResNet-50, show competitive performance, achieving the best results among the \ac{cnn} baselines in both software and hardware encoding scenarios. 

\begin{table*}[t]
\centering
\caption{Performance comparison in YPSNR / VMAF of the considered models on the proposed dataset with VVC encoder. The top result for each encoder type is highlighted in boldface.}
\label{vvc__result}
\resizebox{\textwidth}{!}{%
\begin{tabular}{@{}lllllccc@{}}
\toprule
\multicolumn{2}{l}{Encoder type}        & \multicolumn{6}{c}{VVC software encoding}                                    \\ \cmidrule(l){1-8} 
\multicolumn{2}{l}{Model \textbackslash{} Metric}       & \multicolumn{1}{c}{R2 $\uparrow$} & \multicolumn{1}{c}{SROCC $\uparrow$} & \multicolumn{1}{c}{PLCC $\uparrow$} & Accuracy $\uparrow$       & BD-BR vs EEL $\downarrow$       & BD-BR vs SL $\downarrow$          \\ \midrule
\multirow{4}{*}{\rotatebox[origin=c]{90}{\scriptsize VoD-HandC}} & ExtraTrees        & \textbf{0.5093} / \textbf{0.3379}        & \textbf{0.6696} / \textbf{0.5896}           & 0.7217 / 0.6249          & \textbf{0.9267} / \textbf{0.8030} & \textbf{2.685\%} / \textbf{3.701\%}  & \textbf{-5.184\%} / \textbf{-4.900\%}  \\
& Random Forests    & 0.5057 / 0.2622        & 0.5661 / 0.5251           & \textbf{0.7351} / 0.5654          & 0.9211 / 0.7908 & 3.389\% / 5.051\%  & -4.777\% / -3.487\%  \\
&  XGBoost  & 0.4813 / 0.2096        & 0.5746 / 0.5471           & 0.7065 / 0.5431          & 0.9232 / 0.7968 & 3.448\% / 4.546\%  & -4.692\% / -4.811\%  \\
& LightGBM         & 0.3722 / 0.2354        & 0.5439 / 0.5031           & 0.6147 / 0.5218          & 0.9060 / 0.7989 & 4.009\% / 4.827\%  & -4.596\% / -3.888\%  \\    \addlinespace[0.3ex] \hdashline \addlinespace[0.3ex]
\multirow{4}{*}{\rotatebox[origin=c]{90}{\scriptsize Live-HandC}} & ExtraTrees          & 0.2613 / 0.1144        & 0.5447 / 0.4741           & 0.5611 / 0.2921          & 0.8936 / 0.7792 & 3.714\% / 4.751\%  & -3.890\% / -3.413\%  \\
& Random Forests   & 0.2021 / 0.0511        & 0.5222 / 0.4541           & 0.4700 / 0.2500          & 0.8858 / 0.7662 & 3.863\% / 5.682\%  & -3.764\% / -3.256\% \\
& XGBoost       & 0.2210 / 0.0540        & 0.5351 / 0.4503           & 0.4777 / 0.2705          & 0.8894 / 0.7708 & 3.886\% / 5.395\% & -3.750\% / -3.831\%  \\     
& LightGBM    & 0.1208 / 0.0470         & 0.3641 / 0.3612           & 0.3655 / 0.2606          & 0.8777 / 0.7594 & 5.218\% / 5.533\%  & -2.903\% / -2.984\%  \\ \addlinespace[0.3ex] \hdashline \addlinespace[0.3ex]
\multirow{4}{*}{\rotatebox[origin=c]{90}{\scriptsize Deep features}} & DensNet169         & 0.3998 / 0.3280        & 0.4281 / 0.4924           & 0.6286 / 0.5966          & 0.9018 / 0.7665 & 3.215\% / 4.871\%  & -3.957\% / -3.563\%   \\
& VGG16               & 0.3861 / 0.3066        & 0.4069 / 0.5027           & 0.6288 / 0.5821          & 0.8990 / 0.7676 & 3.606\% / 4.775\%  & -4.012\% / -3.464\%  \\
& ResNet-50         & 0.4220 / 0.3174        & 0.4472 / 0.5712           & 0.6407 / 0.5903          & 0.9144 / 0.7815 & 3.099\% /  4.658\%   & -4.067\% / -4.190\%  \\
& ConvNeXtBase       & 0.3860 / 0.2702        & 0.4373 / 0.5510           & 0.6264 / 0.5293          & 0.9047 / 0.7702 & 3.126\% /  4.801\%   & -3.892\% / -3.392\%  \\  \addlinespace[.55ex] \bottomrule
\end{tabular}%

}\vspace{-5mm}

\end{table*}

Table~\ref{vvc__result} compares the performance of the models considered using the \ac{vvc} software encoder. Among the models considered, the ExtraTrees model, fitted with VoD-HandC features, stands out,  as it achieves the highest performance in terms of \ac{ypsnr} and \ac{vmaf}. It outperforms the static approach with an overall gain of 5.18\%/4.90\% in \ac{ypsnr}/\ac{vmaf} while only incurring a loss of 2.86\%/3.70\% compared to the \ac{eel} method in terms of \ac{bd-br}. Furthermore, the Live-HandC feature-based models consistently perform better than the static approach \ac{sl}, with an average \ac{bd-br} gain of 3.89\%/3.41\% for the best model. In terms of \ac{dnn}-based models, as expected, ResNet-50 outperforms all \ac{cnn} baselines with accuracy in predicting cross-over bitrates of 91\%/78\% in term of \ac{ypsnr}. 

In summary, VoD-HandC feature-based models consistently performed well across different encoding scenarios. The Live-HandC feature-based models also showed superiority over the static approach. Further, \ac{dnn}-based models, particularly ResNet-50, demonstrated competitive performance compared to VoD-HandC feature-based models, which can be further optimized with a more extensive training dataset.

\subsection{Runtime Comparison}
\begin{table}[]
\centering
\caption{Average inference runtime (Seconds) comparison evaluated on 10 UHD videos.}
\label{runtime}
\adjustbox{width=0.45\textwidth}{%
\begin{tabular}{@{}llcccc@{}}
\toprule
\multicolumn{2}{l}{\multirow{2}{*}{Model}} & \multicolumn{2}{c}{Features extraction} & \multicolumn{2}{c}{Prediction} \\ \cmidrule(l){3-6} 
\multicolumn{2}{l}{}                         & CPU                 & GPU               & CPU            & GPU           \\ \midrule
\multirow{4}{*}{\rotatebox[origin=c]{90}{\scriptsize VoD-HandC}}     & ExtraTrees               & 144.754                 & \xmark                    & 0.0011  &  \xmark        \\
                          & Randomforest             & 144.754                 & \xmark                    & 0.0006            & \xmark            \\
                          & XGboost                  & 144.754                 & \xmark                    & 0.0005            & \xmark \\
                          & LightGBM                 & 144.754                 & \xmark                    & 0.0003            & \xmark                \\ \addlinespace[0.3ex] \hdashline \addlinespace[0.3ex]
\multirow{4}{*}{\rotatebox[origin=c]{90}{\scriptsize Live-HandC}}    & ExtraTrees              & 1.086                   & \xmark                    & 0.0013     & \xmark       \\
                          & Randomforest             & 1.086                   & \xmark                    & 0.0008            & \xmark                \\
                          & XGboost                 & 1.086                   & \xmark                    & 0.0006            & \xmark    \\
                          & LightGBM                 & 1.086                   & \xmark                    & 0.0005            & \xmark                \\ \addlinespace[0.3ex] \hdashline \addlinespace[0.3ex]
\multirow{4}{*}{\rotatebox[origin=c]{90}{\scriptsize Deep features}}     & DensNet169                 & 304.877               & 69.041              & 0.1720      & 0.0888      \\
                          & VGG16                    & 473.982               & 67.363              & 0.1403            & 0.0625             \\
                          & ResNet50                 & 273.211               & 63.636              & 0.2157            & 0.1013            \\
                          & ConvexNetBase            & 11729.184             & 146.996             & 0.1541            & 0.0744            \\ \addlinespace[.55ex] \bottomrule
\end{tabular}%
}
\end{table}

Computational complexity is a critical consideration for constructing bitrate ladders in both \ac{vod} and live adaptive streaming applications. We conducted runtime comparisons of the evaluated models on a desktop computer equipped with an Intel® Xeon(R) W-2133 \acs{cpu} @3.60GHz $\times$ 12, 64GB RAM, and a GeForce RTX 2080 Ti \acs{gpu}, running Ubuntu 20.04 LTS. Table~\ref{runtime} presents the average inference time in seconds, measured over ten random \ac{uhd} video sequences from the \ac{avsd} dataset, for both CPU and GPU implementations.

The results indicate that Live-HandC feature-based models achieve remarkably fast feature extraction times, requiring less than 1.1 seconds. Furthermore, the prediction time for these methods is even shorter, ranging from 2 to 13 milliseconds on \acs{cpu}. This demonstrates the suitability of Live-HandC features (\ac{vca} features) for real-time live streaming scenarios where low latency is paramount. Conversely, VoD-HandC feature-based models exhibit a longer feature extraction time of approximately 145 seconds, making them better suited for \ac{vod} streaming. It's worth noting that software optimization of the feature extraction process could potentially reduce time complexity.

Deep learning models also exhibit extended feature extraction times. For example, ResNet50 requires approximately 273 seconds on \acs{cpu} and 63 seconds on \acs{gpu}. However, \ac{dnn} models can be further optimized to reduce inference time through techniques such as pruning and knowledge distillation~\cite{10222892}.

\section{CONCLUSION}
\label{conclusion}
In this paper, we have undertaken a comprehensive review of existing methods for constructing bitrate ladders in adaptive video streaming. These methods are categorized into two main approaches: static and dynamic. Additionally, we have curated an extensive collection of video shots and have developed the largest publicly available dataset of corresponding convex hulls, referred to as \ac{avsd}. These convex hulls were generated by encoding the shots using both hardware and software video encoders, encompassing three standards: \ac{avc}/H.264, \ac{hevc}/H.265, and \ac{vvc}/H.266, across four resolutions and various \ac{qp} values.
Furthermore, we have presented an empirical benchmark study focusing on learning-based methods for convex hull prediction. The experimental results demonstrate that the ExtraTrees Regressor, fitted with VoD-HandC handcrafted features, outperforms other learning-based methods when predicting cross-over bitrates. However, it is worth noting that these features can be computationally expensive and time-consuming, particularly when dealing with \ac{uhd} videos, making them more suitable for \ac{vod} streaming scenarios. Moreover, our complexity analysis reveals that models based on Live-HandC features are the most suitable for live streaming applications. These features strike a balance between inference runtime and prediction performance, making them an efficient choice for such scenarios. Additionally, the promising performances of baseline \ac{cnn} models suggest the significant potential of \ac{dnn} approaches in addressing adaptive streaming challenges. We believe this comprehensive benchmarking study will significantly contribute to and facilitate future research endeavors in the domain of bitrate ladder prediction for adaptive video streaming.


\ifCLASSOPTIONcaptionsoff
  \newpage
\fi

\vspace{-2mm}
\bibliographystyle{IEEEtran}
\bibliography{IEEEabrv,ref_hadi}

\end{document}